\documentclass[reprint,onecolumn,footinbib,longbibliography,superscriptaddress]{revtex4-2}
\usepackage{amsmath}
\usepackage{amsfonts}
\usepackage{amssymb}
\usepackage{times}
\usepackage[dvipdfmx]{graphicx}
\usepackage[usenames,dvipsnames]{xcolor}
\usepackage{bm}
\usepackage{amsthm}
\usepackage[whole]{bxcjkjatype}

\usepackage{physics}
\usepackage{mathrsfs}
\usepackage[hypertexnames=false]{hyperref}  % because we reset counters at the beginning of the Appendix
\usepackage{mathtools}
\usepackage{here}
\usepackage{multirow} % by Fujimoto

\renewcommand{\rm}{\mathrm} % by Fujimoto
\newcommand{\bs}{\boldsymbol} % by Fujimoto
\newcommand{\mc}{\mathcal} % by Fujimoto

\usepackage{comment}

\newcommand{\sn}{\textit{Norm}} % social norm
\newcommand{\snp}{n} % social norm probability
\newcommand{\snptil}{\tilde{n}} % social norm probability with tilde
\newcommand{\ac}{\textit{Action}} % action rule
\newcommand{\acp}{a} % action rule probability
\newcommand{\self}{s}

\begin{document}

\title{Evolutionary stability of cooperation by the leading eight norms in indirect reciprocity\\ under noisy and private assessment}
\author{Yuma Fujimoto}
\email{fujimoto\_yuma@soken.ac.jp}
\affiliation{Research Center for Integrative Evolutionary Science, SOKENDAI (The Graduate University for Advanced Studies). Shonan Village, Hayama, Kanagawa 240-0193, Japan}
\affiliation{Universal Biology Institute (UBI), the University of Tokyo. 7-3-1 Hongo, Bunkyo-ku, Tokyo 113-0033, Japan.}
\affiliation{Department of Evolutionary Studies of Biosystems, SOKENDAI. Shonan Village, Hayama, Kanagawa 240-0193, Japan.}

\author{Hisashi Ohtsuki}
\email{ohtsuki\_hisashi@soken.ac.jp}
\affiliation{Research Center for Integrative Evolutionary Science, SOKENDAI (The Graduate University for Advanced Studies). Shonan Village, Hayama, Kanagawa 240-0193, Japan}

\date{\today}

\begin{abstract}
Indirect reciprocity is a mechanism that explains large-scale cooperation in human societies. In indirect reciprocity, an individual chooses whether or not to cooperate with another based on reputation information, and others evaluate the action as good or bad. Under what evaluation rule (called ``social norm'') cooperation evolves has long been of central interest in the literature. It has been reported that if individuals can share their evaluations (i.e., public reputation), social norms called ``leading eight'' can be evolutionarily stable. On the other hand, when they cannot share their evaluations (i.e., private assessment), the evolutionary stability of cooperation is still in question. To tackle this problem, we create a novel method to analyze the reputation structure in the population under private assessment. Specifically, we characterize each individual by two variables, ``goodness'' (what proportion of the population considers the individual as good) and ``self-reputation'' (whether an individual thinks of him/herself as good or bad), and analyze the stochastic process of how these two variables change over time. We discuss evolutionary stability of each of the leading eight social norms by studying the robustness against invasions of unconditional cooperators and defectors. We identify key pivots in those social norms for establishing a high level of cooperation or stable cooperation against mutants. Our finding gives an insight into how human cooperation is established in a real-world society.
\end{abstract}

\maketitle

\section{Introduction}
Cooperation has been a major topic in biology, psychology, sociology, and economics~\cite{axelrod1984evolution, dugatkin1997cooperation, tomasello2009we, bowles2011cooperative}. Direct reciprocity explains cooperation between two individuals who directly and repeatedly interact with each other~\cite{trivers1971evolution, axelrod1984evolution}. However, cooperation is also seen even in a large-scale society, such as in human societies~\cite{henrich2007humans, tomasello2009we, bowles2011cooperative, boehm2012moral}. This large-scale cooperation is difficult to explain because individuals frequently meet strangers and they do not always interact with the same person. A key to success in large-scale cooperation is social information, such as reputations and gossips. In a real human society, individuals obtain and use the reputations of others to judge whether they cooperate or not. This is a core mechanism of indirect reciprocity, where those who have helped others receive help from a third party through reputations ~\cite{alexander1987biology, nowak1998evolution, nowak2005evolution}. In fact, two-thirds of all human conversations are considered to involve reputations and gossips~\cite{emler1994gossip, dunbar1998grooming, dunbar2004gossip}. Furthermore, many experimental studies support that reputations and gossips contribute to human cooperation~\cite{wedekind2000cooperation, milinski2002reputation, bolton2005cooperation, seinen2006social, sommerfeld2007gossip, barclay2012harnessing, feinberg2014gossip, wu2016reputation, van2016indirect}. 

One of the difficulties in maintaining cooperation by indirect reciprocity concerns errors in choosing actions and assigning reputations. In early studies of indirect reciprocity, the social norm called ``image scoring'' has been discussed~\cite{nowak1998dynamics, nowak1998evolution}. An individual with this social norm assigns a good reputation to those who cooperated and a bad reputation to those who did not. A central question is whether cooperation can be maintained by ``discriminators'', who cooperate with good persons while defect (i.e., not cooperate) with bad persons. In an error-free world, a discriminator cooperates with a good person, and this discriminator obtains a good reputation, which invites cooperation from a third party. However, in a world with errors in actions, a discriminator who accidentally fails to cooperate obtains a bad reputation, which triggers defection from a third party, and this third party obtains a bad reputation, and this triggers another defection, and so on, causing the collapse of cooperation. Therefore, discriminators under the ``image scoring'' social norm cannot maintain cooperation in the presence of errors~\cite{panchanathan2003tale, ohtsuki2004should, ohtsuki2006leading, brandt2005indirect, brandt2006good, ohtsuki2007global, sigmund2010calculus, okada2020review}.

Even under these action and assessment errors, eight social norms have been reported to maintain cooperation in the case of public reputation, where the reputation of each individual is shared among all the individuals~\cite{ohtsuki2004should, ohtsuki2006leading, ohtsuki2007global}. These social norms are called the ``leading eight'' and have the following four properties in common \cite{ohtsuki2006leading}: (i) Maintenance of cooperation; a good person who cooperates with a good person should be evaluated as good, (ii) Identification of defectors; defection with a good person should be evaluated as bad, (iii) Justification of punishment; a good person who defects with a bad person should be evaluated as good, and (iv) Forgiveness: a bad person who cooperates with a good person should be evaluated as good. Here, the above ``image scoring'' satisfies the conditions of (i), (ii), and (iv) but not (iii). Many studies so far have assumed that the reputation is publicly held~\cite{brandt2004logic, pacheco2006stern, suzuki2007evolution, santos2007multi, fu2008reputation, suzuki2008evolutionary, uchida2010competition, ohtsuki2015reputation, santos2016social, santos2016evolution, sasaki2017evolution, santos2018social1, xia2020effect, santos2021complexity, podder2021local} because it is relatively easy to calculate distribution of reputations in the population.

In reality, however, humans do not always agree with others on the reputation of the same individual, so the assumption of public reputation usually does not hold. Instead, they can have different opinions on the same individual (i.e., private assessment). Thus, the reputation structure in the whole population is described by two-dimensional information of who evaluates whom, and how. In a binary reputation case where evaluation is either good or bad, we have a matrix whose elements are good or bad, which is called an image matrix~\cite{uchida2010effect, sigmund2012moral, uchida2013effect, oishi2013group, hilbe2018indirect, fujimoto2022reputation, fujimoto2023evolutionary}. Given that assessment errors occur independently in individuals, this image matrix becomes very complex. This is one of the reasons why the performance of the leading eight social norms under private assessment has so far been investigated based on numerical simulations. It is known that the leading eight social norms cannot maintain cooperation~\cite{hilbe2018indirect} in stochastic processes of invasion and fixation~\cite{traulsen2006stochastic} against those who always cooperate (ALLC) and those who always defect (ALLD). On the other hand, cooperation is maintained if we additionally introduce somewhat special settings such as empathy, generosity, or spatial structure~\cite{brush2018indirect, whitaker2018indirect, radzvilavicius2019evolution, krellner2020putting, quan2020withhold, krellner2021pleasing, schmid2021evolution, quan2022keeping, gu2022reputation, kessinger2023evolution}. Although the recent study~\cite{fujimoto2023evolutionary} has theoretically shown that one of the leading eight norms (called ``L3'' or ``Simple Standing'') can be evolutionarily stable whereas another one (called ``L6'' or ``Stern Judging'') is not, the remaining six of the leading eight norms have yet to be analytically studied. This is partly because these six norms assign reputations to individuals in a more complex manner than the other two~\cite{fujimoto2023evolutionary}, which hindered analytical treatments of those social norms. Therefore, we still do not have a whole picture of which of the leading eight social norms can maintain evolutionarily stable cooperation under private assessment in the presence of errors~\cite{bowles2011cooperative, okada2020review}.

This study investigates whether each of the leading eight norms can sustain evolutionarily stable cooperation or not, by investigating its robustness against invasions of ALLC and ALLD mutants under private assessment with errors. To this end, we develop a new analytical method to analyze stochastic processes of how the image matrix changes, and derive the equilibrium state of the image matrix. Based on this theory, we calculate the payoffs of wild-types and mutants and perform invasion analyses.

\section{A model}
\begin{figure}[h!]
    \centering
    \includegraphics[width=0.55\hsize]{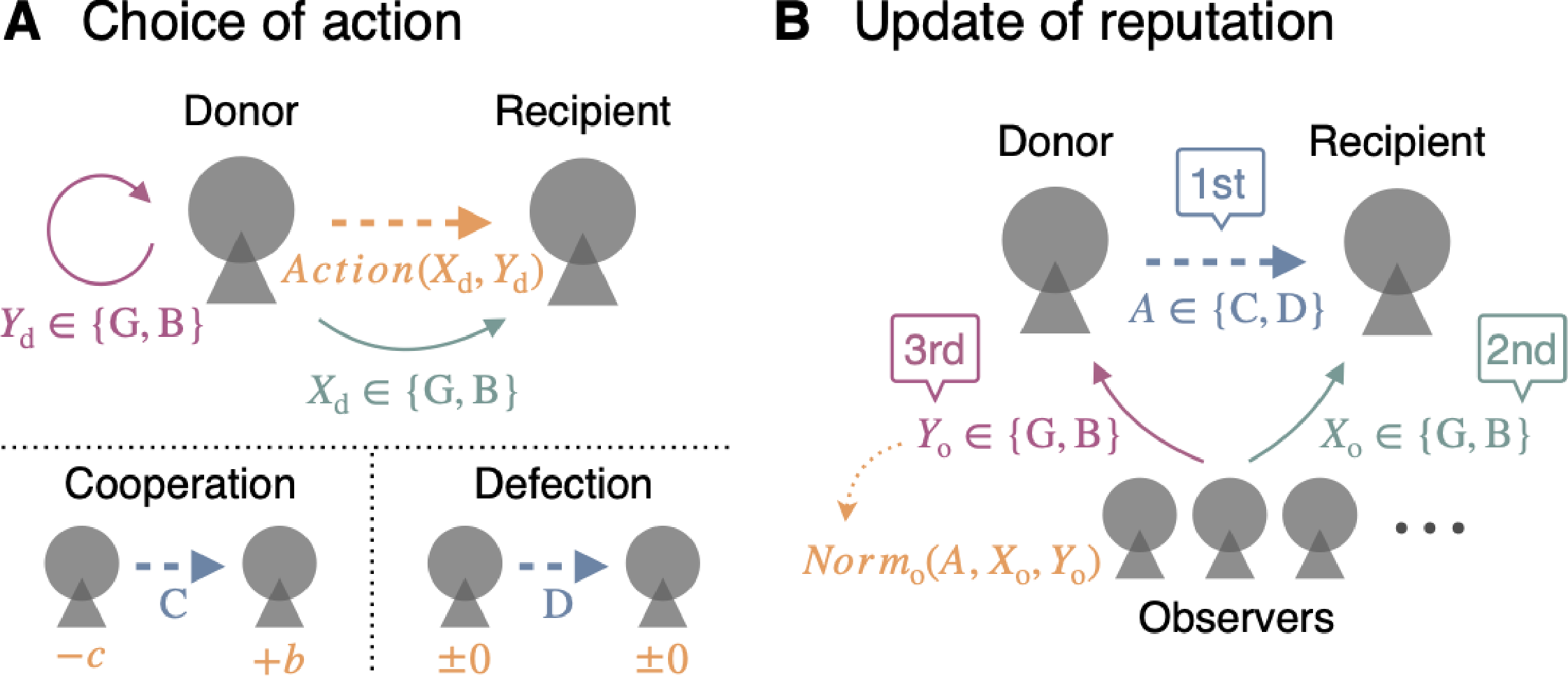}
    \caption{An illustration of how the model of indirect reciprocity proceeds. {\bf A}. The donor chooses the action $\ac(X_{\rm{d}}, Y_{\rm{d}})$ depending on the recipient's ($X_{\rm{d}}$) and donor's ($Y_{\rm{d}}$) reputations in the eyes of the donor. If the donor cooperates ($A=\rm{C}$), the donor pays the cost of $c$, while the recipient gains the benefit of $b$. If the donor defects ($A=\rm{D}$), neither cost nor benefit arises. {\bf B}. Each observer assigns a new reputation $\sn(A, X_{\rm{o}}, Y_{\rm{o}})$ to the donor, depending on the donor's action $A$ (called ``1st-order information''), the recipient's reputation $X_{\rm{o}}$ in the eyes of the observer (called ``2nd-order information''), and the donor's reputation $Y_{\rm{o}}$ in the eyes of the observer (called ``3rd-order information'').
    }
    \label{F00}
\end{figure}

We consider a population of $N$ individuals where a binary reputation of either good ($\rm{G}$) or bad ($\rm{B}$) is given from each individual to each one at any given moment. Reputations can change over time. Every round a donor and a recipient are selected at random from the population. The donor chooses an action of either cooperation ($\rm{C}$) or defection ($\rm{D}$) towards the recipient. If the donor cooperates, he/she pays a cost of $c(>0)$ and gives the recipient a benefit of $b(>c)$. Defection generates neither cost nor benefit. Each player has a rule to choose an action, called ``action rule''. This study assumes that in choosing one's action, the donor considers (1) whether the recipient is good ($X_{\rm{d}}={\rm G}$) or bad ($X_{\rm{d}}={\rm{B}}$) in the eyes of the donor, and (2) whether the donor him/herself is good ($Y_{\rm{d}}={\rm{G}}$) or bad ($Y_{\rm{d}}={\rm B}$) in the eyes of the donor (see Fig.~\ref{F00}A). Thus, the donor's \textit{intended} action is described by a mapping, $\ac(X_{\rm{d}}, Y_{\rm{d}})$, which takes a value of either $\rm{C}$ or $\rm{D}$. Here, we assume that because of an action error, the donor takes the opposite action to the intended one with the probability of $0\le e_1<1/2$. We also assume that the donor is aware of the error when he/she commits it. The situation is formalized as follows. Let $\acp^{X_{\rm{d}}Y_{\rm{d}}}$ be the probability that the donor's \textit{actual} action, denoted by $A$, is cooperation ($\rm{C}$) when the recipient's reputation in the eyes of the donor is $X_{\rm{d}}$ and when the donor's reputation in the eyes of the donor is $Y_{\rm{d}}$. Then we have $\acp^{X_{\rm{d}}Y_{\rm{d}}}=1-e_1$ if $\ac(X_{\rm{d}}, Y_{\rm{d}})={\rm{C}}$, while $\acp^{X_{\rm{d}}Y_{\rm{d}}}=e_1$ if $\ac(X_{\rm{d}},Y_{\rm{d}})={\rm D}$. Thus, the action rule of an individual can be characterized by a $4$-dimensional vector of $\bs{\acp}:=(\acp^{\rm{GG}},\acp^{\rm{GB}},\acp^{\rm{BG}},\acp^{\rm{BB}})$. See Table~\ref{T02} for the list of symbols appearing in this paper.

% Figure 00

\begin{table}[h!]
\centering
\caption{Table of notations}
\begin{tabular}{c|c}
    \hline
    $\rm{C}$, $\rm{D}$ & action: cooperation, defection \\
    $\rm{G}$, $\rm{B}$ & reputation: good, bad \\
    $A\in\{\rm{C},\rm{D}\}$ & variable for action \\
    $X\in\{\rm{G},\rm{B}\}$ & variable for recipient's reputation \\
    $Y\in\{\rm{G},\rm{B}\}$ & variable for donor's reputation \\
    $\rm{d}$, $\rm{o}$ (subscripts of $X$ and $Y$) & in the eyes of donor, observer \\
    $e_1$, $e_2$ & error rates in action and assessment \\
    $b$, $c$ & benefit and cost of cooperation\\
    $N$ & population size \\
    $\sn(A,X,Y)$ (or $\bs{\snp}^{A}$) & social norm \\
    $\ac(X,Y)$ (or $\bs{\acp}$) & action rule \\
    $\self$ & self-reputation \\
    $p$ & goodness\\
    $\phi(p,\self)$ & frequency for self-reputation and goodness \\
    $p_{\rm{M}}$ & goodness of mutants from wild-types \\
    $\phi_{\rm{M}}(p_{\rm{M}})$ & frequency of mutants for goodness \\
    \hline
\end{tabular}
\label{T02}
\end{table}

We assume that every individual in the population observes this interaction as an observer, evaluates the chosen action by the donor, and privately updates the reputation of the donor as good or bad. Every player has his/her own rule to update the donor's reputation, called ``social norm''. We assume that each observer considers the following three pieces of information; (1) whether the donor's \textit{actual} action is cooperation ($A={\rm{C}}$) or defection ($A={\rm{D}}$), (2) whether the recipient is good ($X_{\rm{o}}={\rm{G}}$) or bad ($X_{\rm{o}}={\rm{B}}$) in the eyes of the observer, and (3) whether the donor was good ($Y_{\rm{o}}={\rm{G}}$) or bad ($Y_{\rm{o}}={\rm{B}}$) before the interaction in the eyes of the observer. The reputation that an observer \textit{intends} to assign to the donor is thus represented by the function of $\sn(A, X_{\rm{o}}, Y_{\rm{o}})$, which we call \textit{intended} reputation. Here we also assume that each observer erroneously assigns the opposite reputation to the intended one with probability $0<e_2<1/2$ independently of others. The situation is formalized as follows. We refer to the reputation that is actually assigned to the donor as \textit{actual} reputation, and distinguish it from the \textit{intended} one. We define $\snp^{AX_{\rm{o}}Y_{\rm{o}}}$ as the probability that the \textit{actual} reputation that an observer assigns to the donor is good, when the donor took action $A$, when the recipient's reputation in the eyes of the observer is $X_{\rm{o}}$, and when the donor's previous reputation in the eyes of the observer was $Y_{\rm{o}}$. Then we have $\snp^{AX_{\rm{o}}Y_{\rm{o}}}=1-e_2$ if $\sn(A,X_{\rm o},Y_{\rm{o}})={\rm G}$, while $\snp^{AX_{\rm{o}}Y_{\rm{o}}}=e_2$ if $\sn(A,X_{\rm o},Y_{\rm{o}})={\rm B}$. Each social norm is characterized by two $4$-dimensional vectors $\bs{\snp}^{\rm{C}}:=(\snp^{\rm{CGG}},\snp^{\rm{CGB}},\snp^{\rm{CBG}},\snp^{\rm{CBB}})$ and $\bs{\snp}^{\rm{D}}:=(\snp^{\rm{DGG}},\snp^{\rm{DGB}},\snp^{\rm{DBG}},\snp^{\rm{DBB}})$. When the reputation updates are over, the current donor-recipient pair is resolved, and we repeat the process by sampling a donor and a recipient again. We repeat this process infinitely many times and calculate the expected payoff of each player. In computer simulations for a finite population of size $N$, we assume that $N$ donor-recipient interactions occur in one unit of time.

The leading eight social norms are of particular interest in this study. We label these norms as L1-L8~\cite{sigmund2010calculus}. These norms and their corresponding action rules are described in Table~\ref{T01}. There are several common features in the leading eight, and \citet{ohtsuki2006leading} explained these commonalities as follows: (i) $\sn(\rm{C},\rm{G},\rm{G})=\rm{G}$ and $\ac(\rm{G}, \rm{G})=\rm{G}$ represent ``Maintenance of cooperation''; a good person who cooperates with a good person should be evaluated as good, (ii) $\sn(\rm{D},\rm{G},*)=\rm{B}$ represent ``Identification of defectors''; defection with a good person should be evaluated as bad, (iii) $\sn(\rm{D},\rm{B},\rm{G})=\rm{G}$ and $\ac(\rm{B}, \rm{G})=\rm{D}$ represent ``Punishment \& justification of punishment''; a good person who defects with a bad person should be evaluated as good, and (iv) $\sn(\rm{C},\rm{G},\rm{B})=\rm{G}$ and $\ac(\rm{G}, \rm{B})=\rm{G}$ represent ``Apology and forgiveness''; a bad person who cooperates with a good person should be evaluated as good. On the other hand, the other three pivots in social norms, which are $\sn(\rm{C},\rm{B},\rm{G}), \sn(\rm{C},\rm{B},\rm{B})$ and $\sn(\rm{D},\rm{B},\rm{B})$, were left unspecified; they can be either good or bad. This leads to $2^{3}=8$ combinations, and this is the reason why they are called ``leading eight'' \cite{ohtsuki2006leading}.

% Table 01
\begin{table}
\centering
\caption{Social norms of L1-L8 and their optimal action rules. In the first column, $\sn(\rm{C}, X, Y)$ and $\sn(\rm{D}, X, Y)$ indicate the donor's new reputation assigned by each social norm when the donor cooperates and defects with the recipient, respectively. The column with the heading $\ac(X, Y)$ indicates actions that each donor chooses. In the second row, the first and second alphabets describe the reputations of the recipient ($X$) and the donor ($Y$), respectively.}
\begin{tabular}{c|cccc|cccc|cccc}
    \hline
     & \multicolumn{4}{c|}{$\sn(\rm{C},X,Y)$} & \multicolumn{4}{c|}{$\sn(\rm{D},X,Y)$} & \multicolumn{4}{c}{$\ac(X,Y)$} \\
     $XY$ & GG & GB & BG & BB & GG & GB & BG & BB & GG & GB & BG & BB \\
    \hline
    L1 & G & G & G & G & B & B & G & B & C & C & D & C \\
    L2 & G & G & B & G & B & B & G & B & C & C & D & C \\
    L3 & G & G & G & G & B & B & G & G & C & C & D & D \\
    L4 & G & G & G & B & B & B & G & G & C & C & D & D \\
    L5 & G & G & B & G & B & B & G & G & C & C & D & D \\
    L6 & G & G & B & B & B & B & G & G & C & C & D & D \\
    L7 & G & G & G & B & B & B & G & B & C & C & D & D \\
    L8 & G & G & B & B & B & B & G & B & C & C & D & D \\
    \hline
\end{tabular}
\label{T01}
\end{table}

\section{Analysis of reputation structure}

\subsection{Mathematical framework}
Let us consider a stochastic process describing how the reputations of a chosen donor in the eyes of itself and the others are updated. For a moment, we assume that everyone in the population adopts the same social norm and action rule and that they are one of the leading eight (see Table~\ref{T01}). This assumption will be relaxed later.

We characterize individual's reputations by two variables $(p,\self)$, where $p\in[0,1]$ and $\self\in\{0,1\}$. We define $p\in[0,1]$ as the proportion of the individuals in the population except for the focal one who assigns a good reputation to the focal individual, and call it ``goodness'' of the focal individual. The second variable $\self\in\{0,1\}$ is called ``self-reputation'' of the focal individual; $\self=1$ means that the individual considers him/herself as good, while $\self=0$ if bad. Hereafter we call the pair $(p, \self)$ ``reputation state'' of an individual. For the sake of our later analysis, we introduce vector notations of these variables as $\bs{p}:=(p,1-p)$ and $\bs{\self}:=(\self,1-\self)$.

\begin{figure}
    \centering
    \includegraphics[width=0.8\hsize]{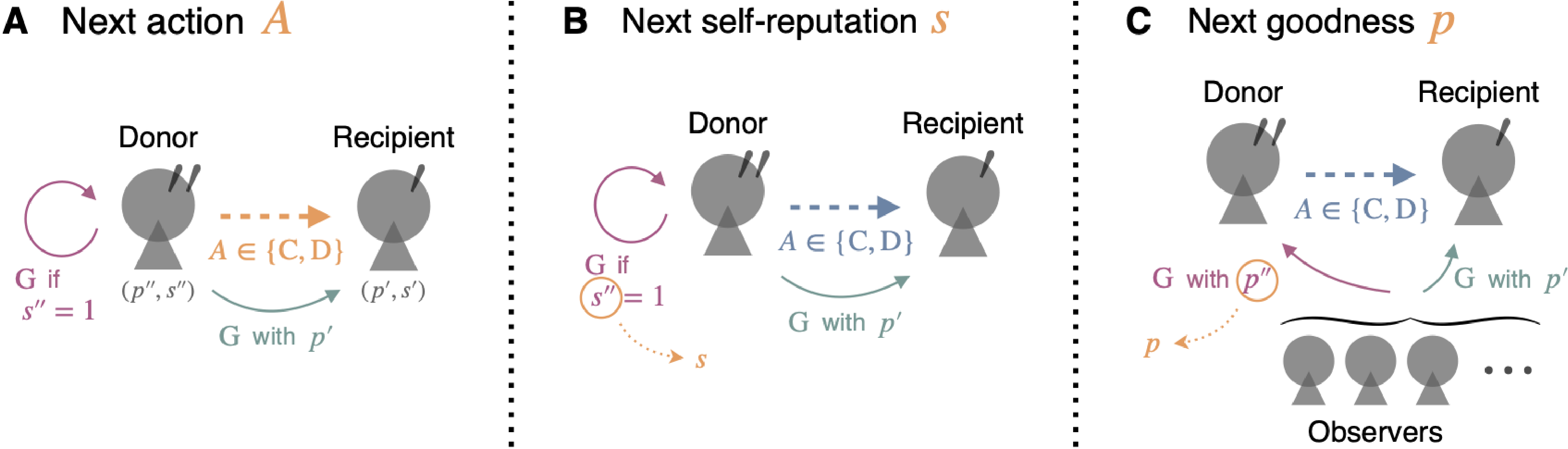}
    \caption{An illustration of the stochastic transition of the donor's reputation state from $(p'',\self'')$ to $(p,\self)$. {\bf A}. First, a donor, whose reputation status is $(p'', \self'')$, either cooperates ($\rm{C}$) or defects ($\rm{D}$) with a recipient, whose reputation status is $(p', \self')$. The action $A\in\{\rm{C}, \rm{D}\}$ depends on the recipient's reputation in the eyes of the donor (given by $p'$) and the donor's self-reputation (given by $\self''$). {\bf B}. The donor updates its self-reputation $\self''$ to $\self$, based on the donor's action $A$, the recipient's reputation in the eyes of the donor (given by $p'$), and the donor's previous self-reputation $\self''$.  {\bf C}. The observers update the donor's reputation in their eyes. The donor's goodness $p''$ is updated to $p$, which depends on the donor's action $A$, the recipient's reputation in the eyes of each observer (given by $p'$), and the donor's previous reputation in the eyes of each observer (given by $p''$).
    }
    \label{F01}
\end{figure}

Now, suppose that the chosen recipient's reputation state is $(p', \self')$ and that the chosen donor's reputation state is  $(p'', \self'')$. We want to calculate the transition probability that the donor's reputation state is updated from $(p'', \self'')$ to $(p, \self)$ after one round of interaction. See Fig.~\ref{F01} for a schematic illustration.

For that purpose, we calculate $h^{AX}$, which is the probability that the donor actually takes action $A \in\{\rm{C},\rm{D}\} $ and actually assigns reputation $X \in\{\rm{G},\rm{B}\}$ to him/herself. It is obtained as
\begin{subequations}\label{hAX}
\begin{align}
    &h^{\rm{CG}}(p',\self'')=(\bs{p}'\otimes\bs{\self}'')\cdot(\bs{\acp}\circ\bs{\snp}^{\rm{C}}),
    \label{hCG}\\
    &h^{\rm{CB}}(p',\self'')=(\bs{p}'\otimes\bs{\self}'')\cdot(\bs{\acp}\circ(\bs{1}-\bs{\snp}^{\rm{C}})),
    \label{hCB}\\
    &h^{\rm{DG}}(p',\self'')=(\bs{p}'\otimes\bs{\self}'')\cdot((\bs{1}-\bs{\acp})\circ\bs{\snp}^{\rm{D}}),
    \label{hDG}\\
    &h^{\rm{DB}}(p',\self'')=(\bs{p}'\otimes\bs{\self}'')\cdot((\bs{1}-\bs{\acp})\circ(\bs{1}-\bs{\snp}^{\rm{D}})).
    \label{hDB}
\end{align}
\end{subequations}
Here, we used tensor product $\bs{p}'\otimes\bs{\self}'' := (p'\self'',p'(1-\self''),(1-p')\self'',(1-p')(1-\self''))$, which generates a vector of probabilities with which the recipient's reputation in the eyes of the donor and the donor's reputation in the eyes of the donor are good-good, good-bad, bad-good, and bad-bad, respectively. The symbol $\circ$ represents the Hadamard product of two $4$-dimensional vectors, which returns a $4$-dimensional vector whose component is a component-wise product of the two original vectors, defined as $(x_{1}, x_{2}, x_{3}, x_{4}) \circ (y_{1}, y_{2}, y_{3}, y_{4}) := (x_{1}y_{1}, x_{2}y_{2}, x_{3}y_{3}, x_{4}y_{4})$. For example, in Eqs.~\eqref{hCG}, the vector $\bs{\acp}$ represents the probabilities with which the donor cooperates with the recipient in each of the four situations above (i.e., good-good, good-bad, bad-good, and bad-bad), and the vector $\bs{\snp}^{\rm{C}}$ represents the probabilities with which the donor assigns a good reputation to him/herself in each situation given that the donor cooperates with the recipient, so their Hadamard product $\bs{\acp}\circ\bs{\snp}^{\rm{C}}$ represents the probabilities with which the donor in each of the four situations above cooperates with the recipient and assigns a good reputation to him/herself. The symbol $\bs{1}$ represents the $4$-dimensional vector with $1$'s in all components. Finally, the symbol $\cdot$ represents the inner product of two vectors.

% Figure 01

Next, we calculate the probability that the donor's goodness changes to $p$ after the interaction given that the donor has chosen action $A \in\{\rm{C},\rm{D}\} $. The vector $\bs{p}'\otimes\bs{p}''=(p'p'',p'(1-p''),(1-p')p'',(1-p')(1-p''))$ represents the proportions of observers who thought that the recipients' reputation and the donor's reputation are good-good, good-bad, bad-good, and bad-bad before the interaction, respectively. Given that the donor has chosen action $A \in\{\rm{C},\rm{D}\}$, each observer independently updates the donor's reputation by using the social norm, which is represented by vector $\bs{\snp}^{A}$, while an assignment error occurs independently. Therefore, for a finite $N$, the value of $p$ is a random variable, and its expectation and variance are calculated as  
\begin{subequations}\label{p_mean_variance}
\begin{align}
    &\rm{E}[p]=(\bs{p}'\otimes\bs{p}'')\cdot\bs{\snp}^{A} \quad \left(=:f^{A}(p',p'')\right),
    \label{fA}\\
    &\rm{Var}[p]=(\bs{p}'\otimes\bs{p}'')\cdot\frac{\bs{\snp}^{A}\circ(\bs{1}-\bs{\snp}^{A})}{N-1}=\frac{e_2(1-e_2)}{N-1} \quad \left(=:V\right),
    \label{variance}
\end{align}
\end{subequations}
which is, for large $N$, rightfully approximated by a Gaussian distribution with mean $\rm{E}[p]$ and variance $\rm{Var}[p]$ by using the central limit theorem. Below, we use this Gaussian approximation. The density function of the Gaussian distribution with mean $\mu$ and variance $\sigma^{2}$ is denoted by $g(x; \mu, \sigma^{2})$:
\begin{equation}
    g(x; \mu, \sigma^{2}) := \frac{1}{\sqrt{2\pi \sigma^{2}}}\exp\left[ - \frac{(x-\mu)^{2}}{2\sigma^{2}}\right].
\end{equation}

Let $\phi(p,\self)$ be a joint probability distribution of individuals whose reputation status is $(p, \self)$. From Eqs.~\eqref{hAX}-\eqref{p_mean_variance}, we can formulate the dynamics of $\phi(p,\self)$. The equilibrium state of $\phi(p,\self)$, denoted by $\phi^{*}(p,\self)$, satisfies
\begin{subequations}\label{phi*}
\begin{align}
    &\phi^{*}(p,1)=\int_0^1\rm{d}p'\sum_{\self'}\int_0^1\rm{d}p''\sum_{\self''}\left\{\sum_{A=\rm{C}, \rm{D}}h^{A\rm{G}}(p',\self'')g(p;f^{A}(p',p''),V)\right\}\phi^{*}(p',\self')\phi^{*}(p'',\self''),
    \label{phi1}\\
    &\phi^{*}(p,0)=\int_0^1\rm{d}p'\sum_{\self'}\int_0^1\rm{d}p''\sum_{\self''}\left\{\sum_{A=\rm{C}, \rm{D}}h^{A\rm{B}}(p',\self'')g(p;f^{A}(p',p''),V)\right\}\phi^{*}(p',\self')\phi^{*}(p'',\self'').
    \label{phi0}
\end{align}
\end{subequations}
On the right-hand side of each equation, the term $\phi^{*}(p',\self')\phi^{*}(p'',\self'')$ represents the probability that the chosen recipient's reputation status is $(p',\self')$ and the donor's one is $(p'',\self'')$ at the equilibrium. The term $h^{A\rm{G}}(p',\self'')g(p;f^{A}(p',p''),V)$ in Eq.~\eqref{phi1} represents the probability the donor takes action $A \in\{\rm{C},\rm{D}\} $ and assigns a good reputation to him/herself, times the probability density that the donor's goodness $p''$ is updated to $p$. Similarly, the term $h^{A\rm{B}}(p',\self'')g(p;f^{A}(p',p''),V)$ in Eq.~\eqref{phi0} corresponds to the case of donor's assigning a bad reputation to itself. Finally, the summations over $\self'$ and $\self''$ and the integrals for $p'$ and $p''$ represent all possible combinations of $(p',\self')$ and $(p'',\self'')$. Note that the Gaussian function $g$ has a positive value in $(-\infty, \infty)$ but we truncate it to $p \in [0, 1]$, so Eq.~\eqref{phi*} is an approximation. As long as the variance of Gaussian function $V$ is small (which is the case if $e_{2}$ is small and/or $N$ is large), however, the error is not so large. As seen in the left-hand sides of Eq.~\eqref{phi*}, the updated distribution should be the same as the original distribution $\phi^{*}$, because $\phi^{*}$ is at the equilibrium.

For an arbitrary function $\mc{X}$, we denote its expected value for the equilibrium distribution as $\bar{\mc{X}}$. For example, the probability with which a recipient receives cooperation in a randomly chosen interaction, denoted by $\bar{h}^{\rm{C}}:=\sum_{X=\rm{G}, \rm{B}}\bar{h}^{\rm{C}X}$, is obtained as
\begin{equation}
    \begin{aligned}
        \bar{h}^{\rm{C}}:=&\int_0^1\rm{d}p'\sum_{\self'}\int_0^1\rm{d}p''\sum_{\self''}\underbrace{\left\{\sum_{X=\rm{G},\rm{B}}h^{\rm{C}X}(p',\self'')\right\}}_{=(\bs{p}'\otimes\bs{\self}'')\cdot\bs{\acp}}\phi^{*}(p',\self')\phi^{*}(p'',\self'')\\
        =& \left[ \underbrace{\left(\int_0^1\rm{d}p'\sum_{\self'}\bs{p}'\phi^{*}(p',\self')\right)}_{=:\bar{\bs{p}}}\otimes\underbrace{\left(\int_0^1\rm{d}p''\sum_{\self''}\bs{\self}''\phi^{*}(p'',\self'')\right)}_{=:\bar{\bs{\self}}} \right] \cdot\bs{\acp}\\
        =&(\bar{\bs{p}}\otimes\bar{\bs{\self}})\cdot\bs{\acp},
    \label{hC}
    \end{aligned}
\end{equation}
where $\bar{\bs{p}}$ and $\bar{\bs{\self}}$ represent the population average of $\bs{p}$ and $\bs{\self}$, respectively.

Eq.~\eqref{phi*} cannot be solved analytically in general. As for the leading eight social norms and action rules (see Table~\ref{T01}), we have previously found ~\cite{fujimoto2022reputation} that, for L3 (its social norm is called ``Simple Standing'') and L6 (its social norm is called ``Stern Judging''), the equilibrium distribution can be analytically derived when it is approximated by a summation of Gaussian distributions, while it cannot for the others, L1, L2, L4, L5, L7, and L8.  This is mainly because social norms in L3 and L6 do not use the previous reputation of the donor in updating donor's reputation. Such social norms are categorized as ``second-order'' norms~\cite{brandt2005indirect, nowak2005evolution, santos2018social1}. In contrast, the other social norms use the reputation of the donor as well, and they are categorized as ``third-order'' norms~\cite{brandt2005indirect, nowak2005evolution, santos2018social1}, which are more complex than second-order ones. 

Consequently, we numerically derive a solution of $\phi^{*}$. To obtain a solution, we replace $\phi^{*}$'s on the right-hand sides of Eq.~\eqref{phi*} with $\phi_{k}$ and those on the left-hand sides with $\phi_{k+1}$ and regard it as a recursion, set $\phi_{0}$ as an arbitrary function (throughout this study, we set it as the uniform distribution function, $\phi_{0}(p,s)=1/2$ for all $p\in[0,1]$ and $s\in\{0,1\}$), sequentially calculate $\phi_{1}, \phi_{2}, \phi_{3}, \cdots$, and regard $\phi_{K}$ as the equilibrium distribution $\phi^{*}$ for some large $K$. During this iteration, we properly normalize each $\phi_{k}$ such that the total probability over $(p,s) \in [0,1] \times \{0, 1\}$ should always be equal to one.  

\subsection{Three different types in the leading eight}
\begin{figure}
    \centering
    \includegraphics[width=0.9\hsize]{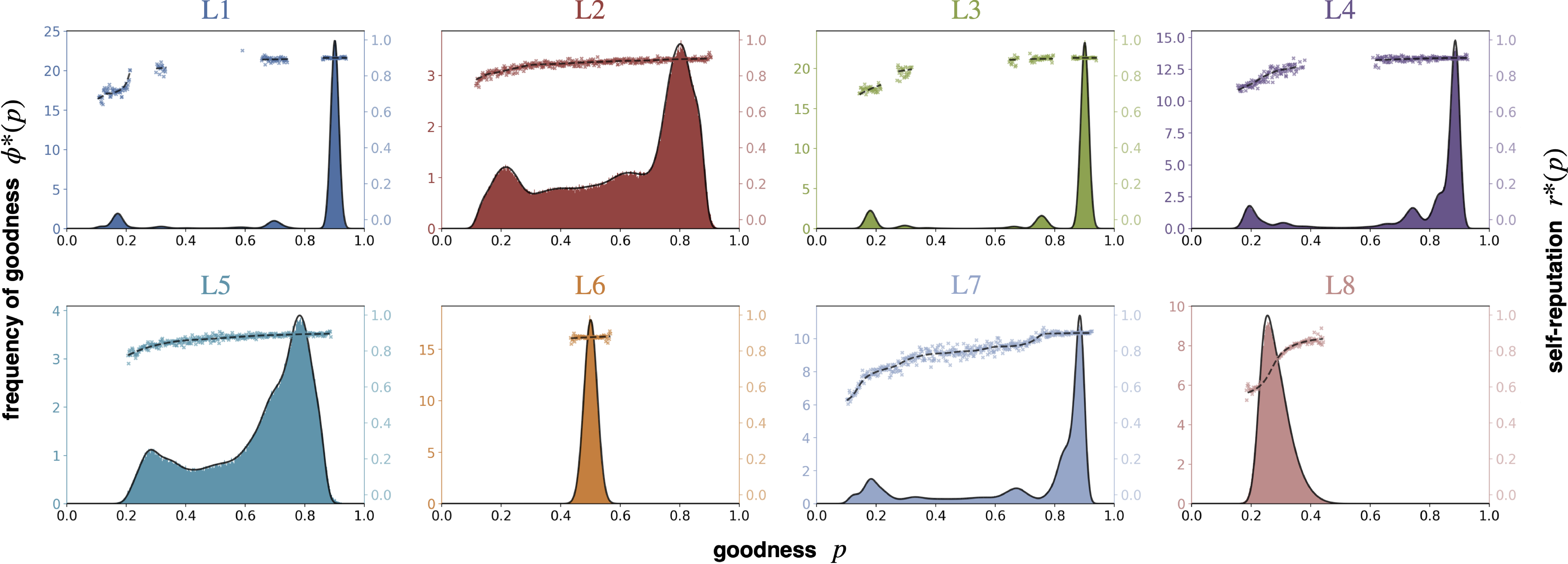}
    \caption{The equilibrium states of $\phi^{*}(p,\self)$ for the leading eight norms (labeled as L1-L8). In each panel, the horizontal axis indicates $p$, goodness of an individual. Colored areas represent $\phi^{*}(p)=\phi^{*}(p,0)+\phi^{*}(p,1)$ in individual-based simulations, that is the marginal distribution of $p$, as indicated by the left axis of the panel. Colored ``x'' markers indicate $r^{*}(p)=\phi^{*}(p,1)/\phi^{*}(p)$ in individual-based simulations, that is the conditional distribution of individuals who assign a good reputation to themselves, given their goodness is $p$, as indicated by the right axis of the panel. We also give the numerical solutions: the black solid lines show numerical solutions of $\phi^{*}(p)$, while the black broken ones show numerical solutions of $r^{*}(p)$. We see nearly a perfect match between individual-based simulations and numerical calculations. The individual-based simulations are based on $(e_1,e_2)=(0.03,0.1)$, $N=501$, and $2000$ samplings from time $51$ to $2050$. The numerical solutions are derived for $N=500$; we stop the iteration when $\|\phi_{K}(p)-\phi_{K-1}(p)\|_{2}<10^{-6}$ is achieved for the first time and regard $\phi_{K}$ as the equilibrium distribution $\phi^{*}$, where $\| \cdot \|_{2}$ is the $L^2$-norm.}
    \label{F02}
\end{figure}

Fig.~\ref{F02} shows the converged distribution by this iterative method for each of the leading eight social norms and its corresponding action rule (see Table~\ref{T01}), which well fits the equilibrium distribution computed from an individual-based simulation. There, the marginal distribution of $p$ defined as $\phi^{*}(p):= \phi^{*}(p,0) + \phi^{*}(p,1)$ and its conditional distribution for $s=1$ defined as $r^{*}(p):=\phi^{*}(p,1)/\phi^{*}(p)$ are simultaneously plotted in the same figure. We remark that the computational complexity of our numerical method is independent of $N$, while that of the individual-based simulation is of order $N^2$. Thus, our numerical method is much more efficient than individual-based simulations for large $N$.

% Figure 02

According to Fig.~\ref{F02}, we can classify the leading eight into three types, as follows.

{\bf (Type-1) L1, L3, L4, and L7}: Those norms are characterized by $\sn(\rm{C}, \rm{B}, \rm{G}) = \rm{G}$ (see Table~\ref{T01}). Under these norms, there is a sharp peak at a very high goodness (i.e.,~around $p=0.9$ in Fig.~\ref{F02}) in the equilibrium distribution, $\phi^{*}(p)$, suggesting that a large majority of individuals in the equilibrium population has a good reputation in the eyes of most observers. The second highest peak is at a very low goodness, although its height is considerably lower than the first one, suggesting that some minority of individuals have a bad reputation in the eyes of most observers. The third highest peak is slightly left to the first peak, the fourth peak is slightly right to the second one, and so on. This behavior, that is, where peaks exist, has already been observed for L3 (``Simple standing'') in our previous study~\cite{fujimoto2023evolutionary}, where we developed its analytical treatment. In short, under these four norms, most of the individuals in the population keep very high goodness in spite of errors.

{\bf (Type-2) L2 and L5}: Those norms are characterized by $\sn(\rm{C}, \rm{B}, \rm{G}) = \rm{B}$ and $\sn(\rm{C}, \rm{B}, \rm{B}) = \rm{G}$ (see Table~\ref{T01}). Under these norms, the equilibrium distribution $\phi^{*}(p)$ looks rather continuous, which is in contrast to Type-1 norms that admit discrete peaks in the equilibrium. The highest peak exists at a high goodness value (i.e.,~around $p=0.8$ in Fig.~\ref{F02}), but this value is not as high as the corresponding highest peak (i.e.,~around $p=0.9$ in Fig.~\ref{F02}) for Type-1 norms. The second highest peak is at a low goodness value. In short, under these two norms, some portion of individuals in the population keep moderately high goodness, while others have either intermediate or low goodness.

{\bf (Type-3) L6 and L8}: Those norms are characterized by $\sn(\rm{C}, \rm{B}, \rm{G}) = \rm{B}$ and $\sn(\rm{C}, \rm{B}, \rm{B}) = \rm{B}$ (see Table~\ref{T01}). Under these two norms, the equilibrium distribution $\phi^{*}(p)$ looks unimodal, and its position is at $p=0.5$ for L6 (``Stern-Judging''), and lower than that for L8. Note that this result for L6 has already been known in previous literature~\cite{uchida2010effect, uchida2013effect, hilbe2018indirect, fujimoto2023evolutionary}. In short, these two norms fail to sustain individuals with a high goodness in the equilibrium population. Since corresponding action norms for L6 and L8 prescribe cooperation with good individuals and defection with bad individuals (see Table~\ref{T01}), it follows that the equilibrium level of cooperation in the population is low, suggesting that indirect reciprocity does not sufficiently work under these norms when assessment is private.

Fig.~\ref{F02} further shows the conditional distribution of self-reputation, $r^{*}(p)$, tends to increase with $p$ in all of the leading eight. This increasing trend means a positive correlation between one's self-reputation and goodness: the one who is evaluated as good by a large proportion of others tends to think of him/herself as good. Self-reputations in L7 and L8 are relatively lower than L1-L6. This is because action rules in L1-L6 always prescribe an action that is evaluated as good under the corresponding social norm, whereas action rules in L7 and L8 do not (see Table~\ref{T01}). More specifically, action rules in L7 and L8 choose $\ac(\rm{B}, \rm{B}) = \rm{D}$ when both the donor and the recipient are bad, but this behavior is evaluated as $\sn(\rm{D}, \rm{B}, \rm{B}) = \rm{B}$ by the corresponding social norms (see Table~\ref{T01}). 

\section{Invasion analysis}
\subsection{Reputations of mutants}
Next, we consider that ALLC or ALLD mutants invade a wild-type population. More specifically, we assume that all the wild-type individuals use one of the leading eight social norms and action rules, and assume that rare mutants, either ALLC or ALLD, attempt to invade the wild-type population. ALLC individuals are those who always intend to cooperate with the recipient. ALLD individuals are those who always intend to defect. We assume that they are susceptible to action errors. On the other hand, they do not need to possess a social norm, because they choose their actions independently of others' reputations. This simple nature of ALLC and ALLD enables us to perform the following invasion analysis.

Because these mutants always intend to choose C or D, their action rules are simple enough. They are represented as $\bs{\acp}_{\rm M} = (\acp_{\rm M}, \acp_{\rm M}, \acp_{\rm M}, \acp_{\rm M})$, where $\acp_{\rm M}=1-e_1$ for ALLC while $\acp_{\rm M}=e_1$ for ALLD. Here and hereafter, symbols with subscript M represent those for mutants, and symbols without subscripts are those for wild-types, unless otherwise specified. We would like to know the payoffs of wild-types and mutants.

Since mutants are rare, the average payoff of wild-types is unaffected by mutants. The probability that a wild-type individual receives cooperation as a recipient remains the same as $\bar{h}^{\rm{C}}$ in Eq.~\eqref{hC}, which is the same as the probability that a wild-type actually performs cooperation as a donor. Therefore, the average payoff of wild-types is
\begin{align}
    &u=(b-c)\bar{h}^{\rm{C}}.
    \label{uwildtype}
\end{align}

As for the payoff of mutants, the probability that a mutant individual cooperates as a donor is $\acp_{\rm M}$. What remains is to derive the probability that a mutant receives cooperation as a recipient, and here we can assume that its donor is a wild-type because the chance of mutant-mutant interactions can be negligibly small.

Since ALLC or ALLD do not use self-reputation, we do not have to consider the self-reputation of these mutants. Thus, we consider only $\phi_{\rm{M}}(p_{\rm{M}})$, which is defined as the probability density function of goodness $p_{\rm{M}}$ of mutants in the eyes of wild-type observers, in the following. Its equilibrium distribution, i.e., $\phi_{\rm{M}}^{*}(p_{\rm{M}})$, should satisfy 
\begin{align}
    \phi_{\rm{M}}^*(p_{\rm{M}})=\int_0^1\rm{d}p'\sum_{\self'}\int_0^1\rm{d}p_{\rm{M}}''\left\{\acp_{\rm M}g(p_{\rm{M}};f^{\rm{C}}(p',p_{\rm{M}}''),V)+(1-\acp_{\rm M})g(p_{\rm{M}};f^{\rm{D}}(p',p_{\rm{M}}''),V)\right\}\phi^{*}(p',\self')\phi_{\rm{M}}^{*}(p_{\rm{M}}''),
    \label{phiM}
\end{align}
where the equation considers how the goodness $p_{\rm{M}}''$ of a mutant donor is updated to $p_{\rm{M}}$ after an interaction with a wild-type recipient whose reputation status is $(p',\self')$. Given that $\phi_{\rm{M}}^{*}(p_{\rm{M}})$ is derived, the probability that a mutant recipient receives cooperation from a wild-type donor, denoted by $\bar{h}_{\rm{M}}^{\rm{C}}$, is calculated as
\begin{equation}
    \begin{aligned}
        \bar{h}_{\rm{M}}^{\rm{C}}=&\int_0^1\rm{d}p_{\rm{M}}'\int_0^1\rm{d}p''\sum_{\self''}\underbrace{\left\{\sum_{X=\rm{G},\rm{B}}h^{\rm{C}X}(p_{\rm{M}}',\self'')\right\}}_{=(\bs{p}'_{\rm{M}}\otimes\bs{\self}'')\cdot\bs{\acp}}\phi_{\rm{M}}^{*}(p_{\rm{M}}')\phi^{*}(p'',\self'')\\
        =& \left[ \underbrace{\left(\int_0^1\rm{d}p'_{\rm{M}}\bs{p}'_{\rm{M}}\phi_{\rm{M}}^{*}(p'_{\rm{M}})\right)}_{=:\bar{\bs{p}}_{\rm{M}}}\otimes\underbrace{\left(\int_0^1\rm{d}p''\sum_{\self''}\bs{\self}''\phi^{*}(p'',\self'')\right)}_{=:\bar{\bs{\self}}} \right] \cdot\bs{\acp}\\
        =&(\bar{\bs{p}}_{\rm{M}}\otimes\bar{\bs{\self}})\cdot\bs{\acp},
        \label{hMC}
    \end{aligned}
\end{equation}
where $\bar{\bs{p}}_{\rm{M}}$ is the average goodness of the mutants in the eyes of wild-types, which is calculated as
\begin{equation}
    \begin{aligned}
        \bar{p}_{\rm{M}}&=\int_0^1\rm{d}p_{\rm{M}}p_{\rm{M}}\phi_{\rm{M}}^{*}(p_{\rm{M}})\\
        &=\int_0^1\rm{d}p'\sum_{\self'}\int_0^1\rm{d}p_{\rm{M}}''\int_0^1\rm{d}p_{\rm{M}}p_{\rm{M}}\left\{\acp_{\rm M}g(p_{\rm{M}};f^{\rm{C}}(p',p_{\rm{M}}''),V)+(1-\acp_{\rm M})g(p_{\rm{M}};f^{\rm{D}}(p',p_{\rm{M}}''),V)\right\}\phi^{*}(p',\self')\phi_{\rm{M}}^{*}(p_{\rm{M}}'')\\
        &=\int_0^1\rm{d}p'\sum_{\self'}\int_0^1\rm{d}p_{\rm{M}}''\left\{\acp_{\rm M}f^{\rm{C}}(p',p_{\rm{M}}'')+(1-\acp_{\rm M})f^{\rm{D}}(p',p_{\rm{M}}'')\right\}\phi^{*}(p',\self')\phi_{\rm{M}}^{*}(p_{\rm{M}}'')\\
        &=\acp_{\rm M}f^{\rm{C}}(\bar{p},\bar{p}_{\rm{M}})+(1-\acp_{\rm M})f^{\rm{D}}(\bar{p},\bar{p}_{\rm{M}}).
        \label{pbarM}
    \end{aligned}
\end{equation}
Here, from the first to the second line we have used Eq.~\eqref{phiM}, from the second to the third line we have used the expectation of Gaussian distribution, and from the third to the fourth line we have taken advantage of the fact that both $f^{\rm{C}}$ and $f^{\rm{D}}$ are multilinear functions. Since the final expression in Eq.~\eqref{pbarM} is linear in $\bar{p}_{\rm{M}}$, we can analytically solve Eq.~\eqref{pbarM} with respect to $\bar{p}_{\rm{M}}$, and it is given as a function of $\bar{p}$. Therefore, we can analytically derive $\bar{h}_{\rm{M}}^{\rm{C}}$ in Eq.~\eqref{hMC}. Finally, the average payoff of mutants is given by
\begin{align}
    &u_{\rm{M}}=b\bar{h}_{\rm{M}}^{\rm{C}}-c\acp_{\rm M}.\label{umutant}
\end{align}
By comparing Eqs.~\eqref{uwildtype} and \eqref{umutant}, we can investigate the invasibility of ALLC and ALLD mutants.

\subsection{Evolutionary stability against ALLC and ALLD}
\begin{figure}
    \centering
    \includegraphics[width=0.9\hsize]{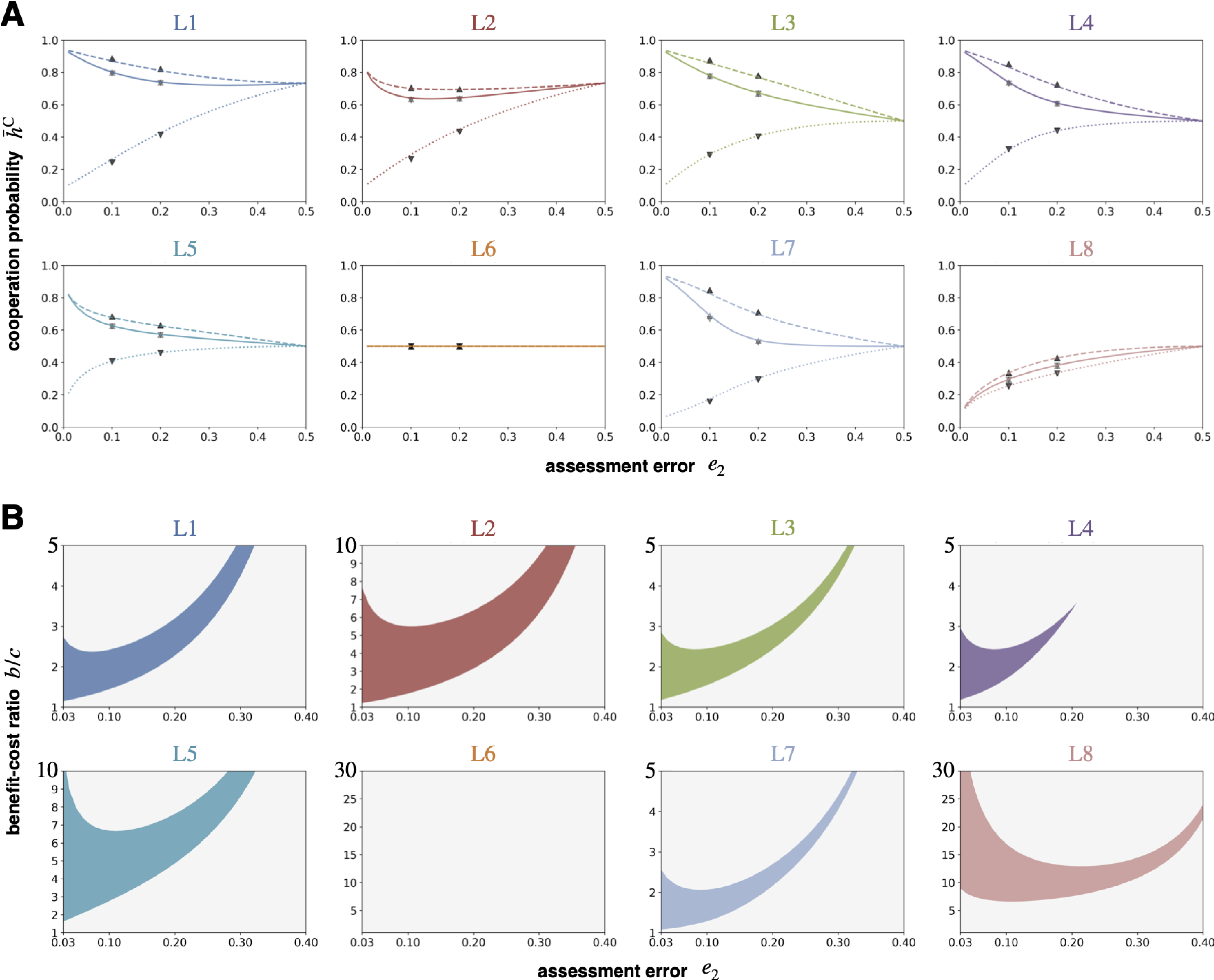}
    \caption{{\bf A}. Cooperation probabilities of the leading eight. Colored lines in each panel show the probabilities that wild-types of each of the leading eight cooperate with wild-types themselves (solid), wild-types cooperate with ALLC mutants (dashed), and wild-types cooperate with ALLD mutants (dotted), respectively. These colored lines are obtained numerically by using Eqs.~\eqref{hC} and \eqref{hMC} for $N=800$ and $e_1=0.03$. Plot marks at $e_2=0.1$ and $e_2=0.2$ show the probabilities of receiving cooperation from wild-type donors calculated from individual-based simulations. Light and dark triangles respectively represent the probabilities that wild-types and mutants receive cooperation when mutants are ALLC, and light and dark inverted triangles respectively represent the probabilities that wild-types and mutants receive cooperation when mutants are ALLD. These plots are from simulations with $N=10000$ and $e_1=0.03$ where the proportion of mutants is set to $0.03$, and averages of $1000$ samples from time steps $51\le t\le 1050$ are shown. {\bf B}. ESS regions of the leading eight against ALLC and ALLD mutants are shown by colored areas. These regions are obtained by the cooperation probabilities plotted in panel A. Note that the scales of y-axis are different between panels.
    }
    \label{F03}
\end{figure}

Using the above equations, we discuss whether or not mutants of ALLC or ALLD can invade wild-types of each of the leading eight. Fig.~\ref{F03}A shows the probability that wild-types cooperate with wild-types (i.e., $\bar{h}^{\rm{C}}$) and two probabilities that wild-types cooperate with ALLC or ALLD mutants (i.e., $\bar{h}_{\rm{M}}^{\rm{C}}$ for $\acp_{\rm M}=1-e_{2}$ and $e_{2}$). Fig.~\ref{F03}B further shows the ESS regions of the leading eight against ALLC and ALLD. The upper bound of the ESS region in each panel corresponds to the invasion condition of ALLC; that is, ALLC can invade the wild-type population if $b/c$ ratio is above that boundary. Similarly, the lower bound of each ESS corresponds to the invasion condition of ALLD; ALLD can invade the wild-type population if $b/c$ ratio is below that boundary.

% Figure 03

\begin{figure}
    \centering
    \includegraphics[width=0.9\hsize]{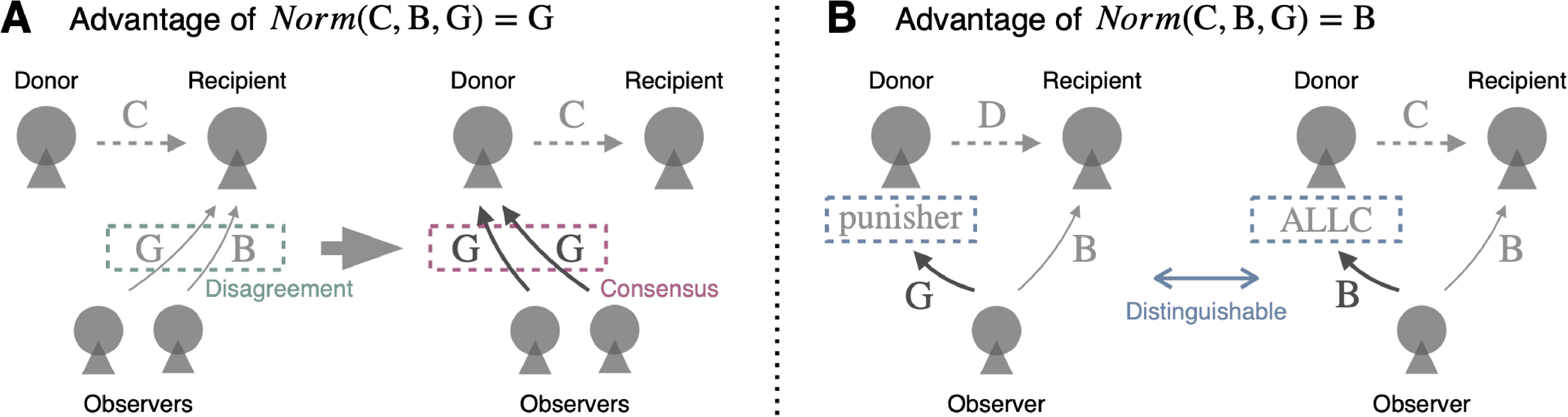}
    \caption{{\bf A}. The advantage of $\sn(\rm{C},\rm{B},\rm{G})=\rm{G}$ lies in that assessment errors cannot spread. In the left panel, both the donor and recipient basically have good reputations. However, only one of two observers (i.e., the right observer) erroneously evaluates the recipient as bad. In the right panel, we show how the observers update the donor's reputation (dark-colored arrows). Despite the disagreement in their opinions on the recipient's reputation, the observers reach a consensus on the donor's reputation because $\sn(\rm{C},\rm{G},\rm{G})=\rm{G}$ for the left observer while $\sn(\rm{C},\rm{B},\rm{G})=\rm{G}$ for the right. {\bf B}. The advantage of $\sn(\rm{C},\rm{B},\rm{G})=\rm{B}$ lies in that unconditional cooperation is detectable. We consider a situation where the recipient is considered as bad in the eyes of all individuals. In the left panel, a good donor who adopts the corresponding action rule of the leading eight chooses defection with the recipient ($\ac(\rm{B}, \rm{G})=\rm{D}$). The observer updates the donor's reputation to a good one (dark-colored arrow) because $\sn(\rm{D},\rm{B},\rm{G})=\rm{G}$. On the other hand, in the right panel the donor is ALLC and unconditionally cooperates with others. The observer updates the ALLC donor's reputation to a bad one (dark-colored arrow) because $\sn(\rm{C},\rm{B},\rm{G})=\rm{B}$. Thus, the leading eight norms satisfying $\sn(\rm{C},\rm{B},\rm{G})=\rm{B}$ can distinguish ALLC from DISC.
    }
    \label{F05}
\end{figure}

{\bf (Type-1) L1, L3, L4, and L7}: These norms can maintain a maximum cooperation level when it is dominant in the population for a small assessment error rate. Indeed, Fig.~\ref{F03}A shows $\bar{h}^{\rm{C}} \to 1-e_1$ ($=0.97$ in this case) in the limit of $e_2\to 0$. Remember that these norms have $\sn(\rm{C},\rm{B},\rm{G})=\rm{G}$ in common (see Table~\ref{T01}). We find that this feature contributes to a high level of cooperation. To see this, imagine two observers of an interaction between a donor and a recipient, and suppose that one of them evaluates the recipient as good, while the other evaluates the recipient as bad possibly due to an assessment error (see Fig.~\ref{F05}A, left). Suppose also that the donor with a good reputation cooperates with the recipient. This cooperation is viewed as ``cooperation with a good recipient by a good donor'' by the former observer, and this observer assigns the reputation $\sn(\rm{C},\rm{G},\rm{G})=\rm{G}$ to the donor. In contrast, this cooperation is viewed as ``cooperation with a bad recipient by a good donor'' by the latter observer, and this observer assigns the reputation $\sn(\rm{C},\rm{B},\rm{G})$, which is $\rm{G}$, to the donor. Therefore, even though observers disagree in their opinions on the recipient, they can reach a consensus on their opinions on the donor (see Fig.~\ref{F05}A, right). This is an intuitive reason why Type-1 norms achieve a very high level of cooperation. The result of study~\cite{fujimoto2023evolutionary} is also explained by the same logic.

In exchange for such maximum cooperation, however, Type-1 norms have relatively narrower ESS regions compared with Type-2. Indeed, Fig.~\ref{F03}B shows that these norms are relatively more vulnerable to the invasion of ALLC for a large $b/c$ than Type-2 norms. This is because Type-1 norms have $\sn(\rm{C},\rm{B},\rm{G})=\rm{G}$ in common, and hence they positively evaluate the unconditional cooperation by ALLC mutants too much. When $e_1=0$ and $e_2\ll 1$, the ESS regions of Type-1 norms are analytically shown to be $1<b/c<2$ (see Appendix~\ref{App_results}).

{\bf (Type-2) L2 and L5}: The cooperation level of these norms is high, but not as high as Type-1. When these norms are dominant in the population and when the assessment error rate is small, the cooperation level is $\bar{h}^{\rm{C}}\simeq 0.8$ in Fig.~\ref{F03}A. Unlike Type-1 norms, these norms have the feature of $\sn(\rm{C},\rm{B},\rm{G})=\rm{B}$ (see Table~\ref{T01}). Imagine again two observers as described in Fig.~\ref{F05}A, who disagree in the opinions on the recipient. Unlike Type-1 norms, when a good donor cooperates with the recipient, the observer who thinks of the recipient as good assigns the reputation $\sn(\rm{C},\rm{G},\rm{G})=\rm{G}$ to the donor, but the observer who thinks of the recipient as bad assigns the reputation $\sn(\rm{C},\rm{B},\rm{G})=\rm{B}$ to the donor. Thus, the disagreement on the recipient's reputation between the two observers will trigger new disagreement, that is the disagreement on the donor's reputation. Bad reputations spread in this way until another common feature of Type-2 norms, $\sn(\rm{C},\rm{B},\rm{B})=\rm{G}$, eventually prevents further spread of bad reputations. This is because when a bad donor cooperates with the recipient, the feature $\sn(\rm{C},\rm{G},\rm{B})=\sn(\rm{C},\rm{B},\rm{B})=\rm{G}$ guarantees that such cooperation is always regarded as good irrespective of how the recipient is viewed from observers. Thus, Type-2 norms can keep their cooperation level high.

At the expense of their cooperation level, Type-2 norms have wider ESS regions than Type-1 norms (see Fig.~\ref{F03}B). This is mainly because Type-2 norms are more resistant to the invasion of ALLC mutants than Type-1. Unlike Type-1 norms, Type-2 observers regard cooperation toward a bad individual as bad ($\sn(\rm{C},\rm{B},\rm{G})=\rm{B}$), hence they occasionally assign bad reputations to unconditional cooperators (i.e., ALLC). Thus, Type-2 norms can selectively assign bad reputations to ALLC mutants but not to residents, leading to wider ESS regions.

{\bf (Type-3) L6 and L8}: Their cooperation level is low ($\bar{h}^{\rm{C}}\le 1/2$). This is because Type-3 norms suffer from disagreements on one's reputation among individuals. Recall that Type-3 norms are characterized by $\sn(\rm{C},\rm{B},\rm{G})=\sn(\rm{C},\rm{B},\rm{B})=\rm{B}$. Since $\sn(\rm{C},\rm{G},\rm{G}) \neq \sn(\rm{C},\rm{B},\rm{G})$, by the same logic as Type-2 norms, disagreement between observers about a recipient propagates to that about a donor. Furthermore, the feature $\sn(\rm{C},\rm{G},\rm{B}) \neq \sn(\rm{C},\rm{B},\rm{B})$ further worsens such disagreements. Due to the accumulation of assessment errors, under Type-3 norms individuals fail to synchronize their evaluations of the same individual at an equilibrium, as suggested by the unimodal peak at an intermediate $p$ value in Fig.~\ref{F02}, leading to a low cooperation level.

Furthermore, Type-3 norms are fragile against the invasion of ALLD mutants. Recall that individuals under Type-3 norms have only intermediate levels of goodness (see Fig.~\ref{F02}). Since $\sn(\rm{D},\rm{B},\rm{G})=\rm{G}$, these norms consider ALLD mutants as good to some extent. Thus, their cooperation levels toward themselves and ALLD mutants can differ little. Indeed, Fig.~\ref{F03}A shows that L6 cannot distinguish themselves and ALLD, while L8 barely does. Fig.~\ref{F03}B also shows that L6 is always invaded by ALLD, while L8 is invaded by ALLD unless the benefit-cost ratio is unrealistically high $(b/c > 5\sim 10)$.
\color{black}

% Figure 04

\section{Conclusion and discussion}
The leading eight norms have been known to maintain cooperation under public reputations in indirect reciprocity~\cite{ohtsuki2004should, ohtsuki2006leading}. Whether these norms can maintain cooperation even under private assessments (where all individuals independently evaluate others), however, has been largely an open question (but see \cite{hilbe2018indirect}). In the present paper, we have extended our previous analytical framework for studying second-order social norms \cite{fujimoto2022reputation, fujimoto2023evolutionary} and developed a novel methodology that enables us to study third-order social norms. We have revealed that the leading eight norms can be classified into three types based on the shape of their reputation structure and the degree of cooperation. We have also discussed the invasibility of ALLC and ALLD mutants and found that these three types of social norms are different in their resistance to invasion by those mutants. In addition, we have provided intuitive explanations of where this difference originates from.

Specifically, this study has shown that Type-1 norms have a maximal cooperation level but are relatively weak to the invasion of ALLC mutants, leading to relatively narrow ESS regions. On the other hand, Type-2 norms have a lower cooperation level than Type-1 norms, but their ESS regions are relatively wider due to their resistance to ALLC mutants. Thus, there is a trade-off between their cooperation levels and resistance to ALLC mutants. We identified that $\sn(\rm{C},\rm{B},\rm{G})$ is a key pivot and that it is the source of this trade-off. The pivot $\sn(\rm{C},\rm{B},\rm{G})$ represents how to evaluate cooperation by a good donor toward a bad recipient from the viewpoint of observers. If this evaluation is good (i.e.,~Type-1 norms), then an advantage arises that the spread of disagreements among individuals is avoided, with a disadvantage that individuals show generosity toward unconditional cooperation by ALLC mutants. On the other hand, if this evaluation is bad (i.e.,~Type-2 and Type-3 norms), the opposite effects appear. Whether the further spread of disagreements occurs or not is determined by another key pivot, $\sn(\rm{C},\rm{B},\rm{B})$. If this evaluation is good (i.e.,~Type-2 norms), cooperation is sustained, whereas if this evaluation is bad (i.e.,~Type-3 norms), sustaining cooperation is difficult.

Fujimoto \& Ohtsuki~\cite{fujimoto2023evolutionary} specified parameter regions where each of the second-order norms is not invaded by the other norms. The leading eight norms contain two second-order norms, L3 and L6 (called $S_{03}$, $S_{06}$ in their study). Our finding here that L3 can be evolutionarily stable but L6 cannot is consistent with their result (although detailed settings are different between these studies: for example, our study assumes ALLC and ALLD as potential mutants, while their study considered ALLG and ALLB mutants, which always assign good or bad reputations to everyone, and they studied several other mutants as well). In this sense, the current study is an extension of \citet{fujimoto2023evolutionary}; we have studied the other six leading eight social norms as well, classified them, and investigated their property.

An example of expected future studies is to extend the analysis to more general situations. First, the current study considered only ALLC and ALLD as potential invaders. It would be interesting to study other types of mutants and see the robustness of Type-1 and Type-2 norms in the leading eight. This includes studying mutual invasibility among the leading eight norms. Second, the leading eight norms are just eight norms out of $2^{8}=256$ possible third-order norms, so exhaustively studying the property of each of the 256 third-order norms is important in order to clarify the role of third-order norms under private assessment. Third, studying higher-order norms, which use more information in assigning a new reputation such as the previous reputation of the recipient (called fourth-order information~\cite{santos2018social1, santos2021complexity}), is an interesting direction of extension. In particular, how the norm complexity~\cite{santos2018social1, santos2021complexity} (concerning the order of norms) is related to its ability to maintain cooperation is an open question under noisy and private assessment. Fourth, we have assumed for simplicity that everyone in the population observes every interaction in the population. However, it is a highly idealized situation, and it would be more realistic to consider the situation of partial observation, in which only a fraction (say, fraction $0 < \theta <1$) of individuals can observe a given interaction and update reputations. Fifth, it could be interesting to investigate the reputation structure of the population under non-binary reputation, such as ternary reputation~\cite{tanabe2013indirect, murase2022social}, where each individual can be labeled as good, bad, or neutral.

In conclusion, we have successfully classified the leading eight norms into three types according to their performance under noisy and private assessment. From the perspective of the cooperation level and resistance to ALLC and ALLD mutants, we find that Type-1 and Type-2 norms are promising. In particular, the pivot $\sn(\rm{C},\rm{B},\rm{G})$ in the social norm determines the trade-off between cooperation level and resistance to ALLC. It is eagerly awaited to experimentally examine whether $\sn(\rm{C},\rm{B},\rm{G})$ is evaluated as good or bad in real societies and how this pivot actually contributes to sustaining a high level of cooperation by indirect reciprocity.

\section*{Acknowledgments}
Y.F. acknowledges the support by JSPS KAKENHI Grant Number JP21J01393. H.O. acknowledges the support by JSPS KAKENHI Grant Numbers JP19H04431 and JP23K03211.

\section*{Author Contributions}
Y.F. and H.O. designed research, calculated results numerically and analytically, and wrote the paper.

\section*{Availability of Data and Materials}
The datasets used and/or analyzed during the current study available from the corresponding author on reasonable request.

%\bibliographystyle{apsrev4-2}
%\bibliography{biblio.bib}


\newpage
\appendix

\renewcommand{\thefigure}{A\arabic{figure}}
\setcounter{figure}{0}

\begin{center}
{\Large {\bf Appendix}}
\end{center}

\section{Results for $e_1=0$} \label{App_results}
\subsection{Simulation results}
\begin{figure}[h!]
    \centering
    \includegraphics[width=0.9\hsize]{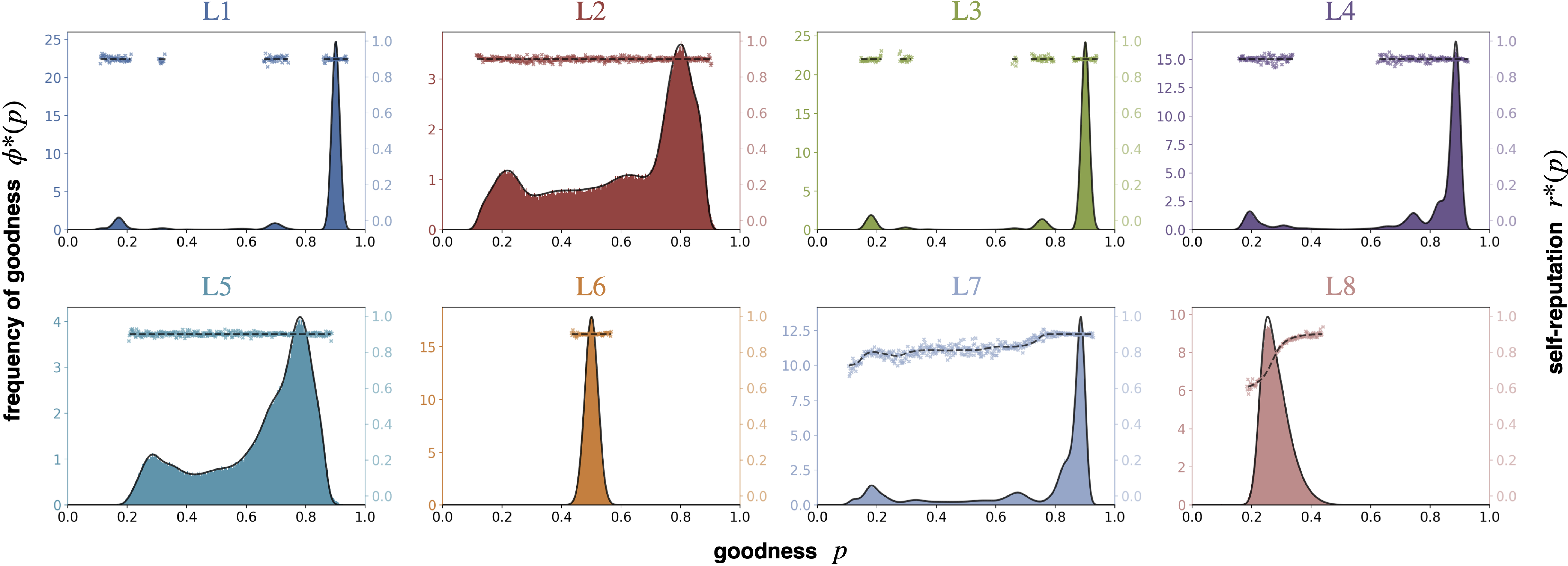}
    \caption{The equilibrium states of $\phi^*(p,\self)$. All the panels are output based on the same parameters as Fig.~\ref{F02} except for $e_1=0$.
    }
    \label{FS01}
\end{figure}

In the main text, we have assumed $e_1=0.03$. In order to provide theoretical insights into the differences among the leading eight norms, this section focuses on the case of no action error, $e_1=0$. First, Fig.~\ref{FS01} shows the reputation structure of the leading eight norms in the same way as Fig.~\ref{F02} but for $e_1=0$. This figure shows that the leading eight norms can be again categorized into three types in their reputation structure, namely, L1, L3, L4, and L7 are in Type-1, L2 and L5 are in Type-2, and L6 and L8 are in Type-3. The only qualitative difference is that the conditional distribution for $p$ given that $s=1$ is constant, $r^*(p)=1-e_2$, for L1-L6. This is because a donor with L1-L6 is always able to take actions that are evaluated as good in its own eyes (because $e_{1}=0$ and $\sn(\ac(X,Y), X, Y )=\rm{G}$ for any $X$ and $Y$; see Table~\ref{T01}) and actually evaluates itself as good unless an assessment error occurs. On the other hand, a donor with L7-L8 inevitably chooses actions that are evaluated as bad in its own eyes when both the recipient and donor are bad in its own eyes (i.e, $\sn(\ac(B,B), B, B)=\rm{B}$); see Table~\ref{T01}). Thus, $r^{*}(p)=1-e_2$ is not satisfied for L7-L8.
% Figure S01

\begin{figure}[htbp]
    \centering
    \includegraphics[width=0.9\hsize]{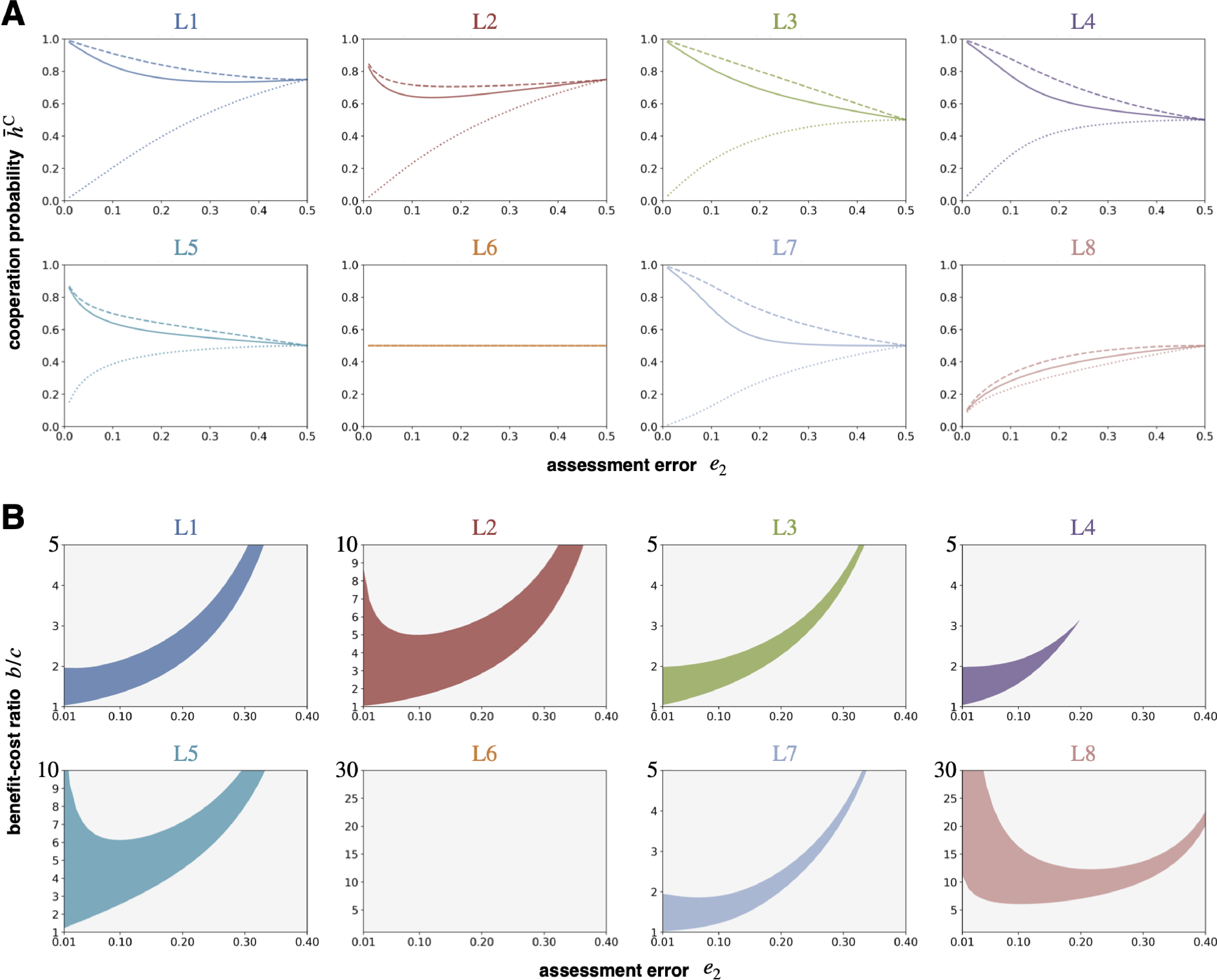}
    \caption{{\bf A}. Cooperation probabilities of the leading eight norms. {\bf B}. ESS regions of the leading eight norms. Numerical calculations are based on the same parameters as Fig.~\ref{F03} except for $e_1=0$.}
    \label{FS02}
\end{figure}

Fig.~\ref{FS02} shows the cooperation probabilities and ESS regions of the leading eight norms for $e_1=0$ for the sake of comparison with Fig.~\ref{F03}. From Fig.~\ref{FS02}, we see that the results of Fig.~\ref{F03} look robust against the change of the action error rate from $e_{1}=0.03$ to $e_{1}=0$. However, panel B in Fig.~\ref{FS02} for Type-1 norms suggests that the ESS condition in the limit of $e_2\to 0$ for those norms is $1<b/c<2$. Below, we will analytically prove this result.
% Figure S02

\subsection{Analytical results}
\subsubsection{Equilibrium distribution}
Let us assume $e_1=0$ and consider the limit, $N\to\infty$. Our temporal goal is to derive the marginal distribution, $\phi^{*}(p) := \phi^{*}(p,1) + \phi^{*}(p,0)$, at the equilibrium when there are no mutants. 

First, we consider the quantity $\sum_{s''}\sum_{X}h^{\rm{C}X}(p',s'')\phi^*(p'',s'')$; i.e., the probability that a randomly chosen donor has goodness $p''$ and it cooperates with the recipient whose goodness is $p'$ at the equilibrium. We notice that the following properties are satisfied:
\begin{itemize}
\item For L1-L2, since we assume $e_1=0$, donors always choose the action evaluated as good by themselves unless an assessment error occurs. Thus, $r^*(p)=1-e_2$ holds for all $p$, and therefore 
\begin{equation}
    \phi^{*}(p'', s'') =
    \begin{cases}
        \phi^{*}(p'') \cdot (1-e_{2}) & \text{if $s''=1$} \\
        \phi^{*}(p'') \cdot e_{2} & \text{if $s''=0$}
    \end{cases}
\end{equation}
holds. Hence we obtain 
\begin{equation}
    \sum_{s''}\sum_{X}h^{\rm{C}X}(p',s'')\phi^*(p'',s'')=\phi^*(p'')\sum_{X}\{ \underbrace{h^{\rm{C}X}(p',1)}_{=p'}(1-e_2)+\underbrace{h^{\rm{C}X}(p',0)}_{=1} e_2 \}=\phi^*(p'') \{p'+(1-p') e_2\}.
\end{equation}
\item For L3-L8, donors do not change their actions depending on the self-reputation. So, we can obtain $\sum_{X}h^{\rm{C}X}(p',s'')=p'$. Thus, we obtain 
\begin{equation}
    \sum_{s''}\sum_{X}h^{\rm{C}X}(p',s'')\phi^*(p'',s'')=p'\sum_{s''}\phi^*(p'',s'')=\phi^*(p'')p'.
\end{equation}
\end{itemize}
To summarize these properties, we define $h^{\rm{C}}(p')$ as follows;
\begin{equation}
\begin{aligned}
    \sum_{s''}\sum_{X}h^{\rm{C}X}(p',s'')\phi^*(p'',s'')&=\phi^*(p'')\left\{\begin{array}{ll}
        p'+(1-p')e_2 & (\rm{L1}-\rm{L2}) \\
        p' & (\rm{L3}-\rm{L8})
    \end{array}\right.\\
    &=\phi^*(p'')\underbrace{(p'+(1-p')e_2\acp^{\rm{BB}})}_{=:h^{\rm{C}}(p')},
    \label{hC_p}
\end{aligned}
\end{equation}
because $\acp^{\rm{BB}}=1$ for L1-L2 and $\acp^{\rm{BB}}=0$ for L3-L6. Note that $h^{\rm{C}}(p')$ represents the probability that a random donor who faces the recipient of goodness $p'$ cooperates with him/her at the equilibrium. We also define $h^{\rm{D}}(p'):=1-h^{\rm{C}}(p')$. With these, the equilibrium marginal distribution of goodness is given, from Eq.~\eqref{phi*}, by
\begin{equation}
\begin{aligned}
    \phi^*(p)&=\sum_{s}\phi^*(p,s)\\
    &=\int_{0}^{1}dp'\sum_{s'}\int_{0}^{1}dp''\sum_{s''}\sum_{A}\sum_{X}h^{AX}(p',s'')\delta(p-f^{A}(p',p''))\phi^*(p',s')\phi^*(p'',s'')\\
    &=\int_{0}^{1}dp'\int_{0}^{1}dp''\sum_{A}\underbrace{\sum_{s''}\sum_{X}h^{AX}(p',s'')\phi^*(p'',s'')}_{=h^{A}(p')\phi^*(p'')}\delta(p-f^{A}(p',p''))\phi^*(p')\\
    &=\int_{0}^{1}dp'\int_{0}^{1}dp''\sum_{A}h^{A}(p')\delta(p-f^{A}(p',p''))\phi^*(p')\phi^*(p'').
    \label{phi*_p}
\end{aligned}
\end{equation}
Here, we used the fact that in the limit of $N\to\infty$ the Gaussian $g(p;f^{A}(p',p''),V)$ converges to the Dirac delta function $\delta(p-f^{A}(p',p''))$.

We now consider the case of $e_2\ll 1$. In order to calculate the ESS condition of Type-1 norms (L1, L3, L4, and L7), we need to find the solution to Eq.~\eqref{phi*_p} up to the order of $e_2$. In other words, we neglect all the terms of $O(e_{2}^{2})$ and higher in the following calculations. Guessing from Fig.~\ref{FS01}, we heuristically seek the solution in the following form:
\begin{subequations}\label{candidate_form}
\begin{equation}\label{canditate}
    \phi^{*}(p) = \sum_{i=1}^{m_1} \kappa_{i} \delta_{1-k_{i} e_{2}} (p) + \sum_{i=1}^{m_2} \lambda_{i} \delta_{l_{i} e_{2}} (p). 
\end{equation}
Here, $\delta_{x}(p) := \delta(p-x)$ is the Dirac delta function that has a unit mass at $p=x$, and $\kappa$'s and $\lambda$'s satisfy
\begin{equation}
    \kappa_{i} = \text{$O(1)$ or $O(e_{2})$}, \quad \lambda_{i} = O(e_{2}), \quad \sum_{i=1}^{m_1} \kappa_{i} + \sum_{i=1}^{m_2} \lambda_{i} =1
\end{equation}
\end{subequations}
In words, Eq.~\eqref{candidate_form} says that the equilibrium distribution $\phi^{*}$ is given as a finite sum of masses $\kappa_{i}$ at positions $p=1-k_{i} e_{2}$ and masses $\lambda_{i}$ at positions $p=l_{i} e_{2}$, where masses $\kappa_{i}$ are either $O(1)$ or $O(e_{2})$ and masses $\lambda_{j}$ are $O(e_{2})$. 

In fact, calculations in Appendix~\ref{App_detailed} show that we can find the solution to Eq.~\eqref{phi*_p} in the form of Eq.~\eqref{candidate_form} for each of the Type-1 norms (L1, L3, L4, and L7) separately as
\begin{equation}\label{phi*_solution}
    \begin{split}
        \phi^{*}_{\mathrm{L1}}(p) &= (1-2e_{2}) \cdot \delta_{1-e_{2}}(p) + e_{2} \cdot \delta_{1-4e_{2}}(p) + e_{2} \cdot \delta_{2e_{2}}(p) \\
        \phi^{*}_{\mathrm{L3}}(p) &= (1-2e_{2}) \cdot \delta_{1-e_{2}}(p)  + e_{2} \cdot \delta_{1-3e_{2}}(p) + e_{2} \cdot \delta_{2e_{2}}(p) \\
        \phi^{*}_{\mathrm{L4}}(p) &= (1-3e_{2}) \cdot \delta_{1-e_{2}}(p)  + e_{2} \cdot \delta_{1-2e_{2}}(p) + e_{2} \cdot \delta_{1-3e_{2}}(p) + e_{2} \cdot \delta_{2e_{2}}(p) \\
        \phi^{*}_{\mathrm{L7}}(p) &= (1-3e_{2}) \cdot \delta_{1-e_{2}}(p)  + e_{2} \cdot \delta_{1-2e_{2}}(p) + e_{2} \cdot \delta_{1-4e_{2}}(p) + e_{2} \cdot \delta_{2e_{2}}(p). 
    \end{split}
\end{equation}
Note that these solutions are correct up to $O(e_{2})$. We also remark that with the same methodology, we cannot derive the equilibrium distribution for Type-2 norms (L2 and L5), because we eventually find that sums in Eq.~\eqref{canditate} cannot be finite but require infinite sums, which is consistent with our observation that the equilibrium distribution $\phi^{*}(p)$ for Type-2 norms look continuous, not discrete.

\subsubsection{A calculation of payoffs}
With the equilibrium distribution given by Eq.~\eqref{phi*_solution} we can calculate various quantities for studying the invasion condition of ALLC and ALLD mutants. In the following calculation we use the facts that Type-1 norms and the corresponding action rules are given by $\bs{\snp}^{\rm{C}}=(1-e_2,1-e_2,1-e_2,\snp^{\rm{CBB}})$, $\bs{\snp}^{\rm{D}}=(e_2,e_2,1-e_2,\snp^{\rm{DBB}})$, and $\bs{\acp}=(1,1,0,\acp^{\rm{BB}})$.

{\bf Expected payoff of wild-types}: From Eq.~\eqref{phi*_solution}, the average goodness $\bar{p}$ of Type-1 wild-types is calculated as
\begin{equation}
    \bar{p} = \int_{0}^{1} p \phi^{*}(p) \mathrm{d}p =1-2e_2+o(e_2)
\end{equation}
for all Type-1 norms. The probability of giving/receiving cooperation $\bar{h}^{\rm{C}}$ in the monomorphic population of residents at the equilibrium is
\begin{equation}
    \bar{h}^{\rm{C}}=h^{\rm{C}}(\bar{p})=\bar{p}+(1-\bar{p})e_2\acp^{\rm{BB}}
    =1-2e_2+o(e_2)
\end{equation}
for all Type-1 norms.
Therefore, from Eq.~\eqref{uwildtype}, the expected payoff of wild-types is
\begin{equation}\label{uType1}
    u=(1-2e_2)(b-c)+o(e_2)
\end{equation}
for all Type-1 norms.

{\bf Expected payoff of ALLC mutants}: Consider rare ALLC mutants invading one of the Type-1 norms. By putting $a_{\rm{M}}=1$ in Eq.~\eqref{pbarM}, the average goodness of these mutants is
\begin{equation}
    \bar{p}_{\rm{M}}
    =f^{\rm{C}}(\bar{p},\bar{p}_{\rm{M}})
    = 
    \begin{cases}
        1-e_{2} & \text{(for L1 and L3)} \\
        1-e_{2} - (1-2e_{2})(1-\bar{p})(1-\bar{p}_{\rm{M}}) & \text{(for L4 and L7)}
    \end{cases}
\end{equation}
For L4 and L7, this is further calculated as
\begin{equation}
    \bar{p}_{\rm{M}}=\frac{1-e_2-(1-2e_{2})(1-\bar{p})}{1-(1-2e_{2})(1-\bar{p})} = \frac{1-e_2-(1-2e_{2})2e_{2}}{1-(1-2e_{2})2e_{2}} + o(e_{2}) = 1 - e_{2} + o(e_{2}).
\end{equation}
From Eq.~\eqref{hC}, the expected probability that a wild-type cooperates with an ALLC mutant is
\begin{equation}
    \bar{h}_{\rm{M}}^{\rm{C}}=h^{\rm{C}}(\bar{p}_{\rm{M}})=\bar{p}_{\rm{M}}+(1-\bar{p}_{\rm{M}})e_2\acp^{\rm{BB}} =1-e_2+o(e_2)
\end{equation}
for all Type-1 wild-types.
By putting $a_{\rm{M}}=1$ in Eq.~\eqref{umutant} we obtain the expected payoff of ALLC mutants as
\begin{equation}\label{uALLCforType1}
    u_{\rm{M}}=(1-e_2)b-c+o(e_2)
\end{equation}
for all Type-1 wild-types.
Thus, by comparing Eq.~\eqref{uType1} and Eq.~\eqref{uALLCforType1}, wild-types are resistant to ALLC mutants if and only if
\begin{align}
    b/c< 2
\end{align}
in the limit of $e_{2} \to 0$ for all Type-1 wild-types.

{\bf Expected payoff of ALLD mutants}: Consider rare ALLD mutants invading one of the Type-1 norms. By putting $a_{\rm{M}}=0$ in Eq.~\eqref{pbarM}, the average goodness of these mutants is
\begin{equation}
\begin{aligned}
    \bar{p}_{\rm{M}}
    &=f^{\rm{D}}(\bar{p},\bar{p}_{\rm{M}})\\
    &=e_2\bar{p}+(1-e_2)(1-\bar{p})\bar{p}_{\rm{M}}+\snp^{\rm{DBB}}(1-\bar{p})(1-\bar{p}_{\rm{M}})\\
    \Longleftrightarrow \quad \bar{p}_{\rm{M}}
    &=\frac{e_2 \bar{p} + \snp^{\rm{DBB}}(1-\bar{p})}{1-(1-e_2)(1-\bar{p}) + \snp^{\rm{DBB}}(1-\bar{p})}\\
    &=\frac{e_2(1+2\snp^{\rm{DBB}})}{1-2e_2(1-\snp^{\rm{DBB}})} + o(e_{2}).
\end{aligned}
\end{equation}
Since $\snp^{\rm{DBB}} = e_{2}$ for L1 and L7, and $\snp^{\rm{DBB}} = 1-e_{2}$ for L3 and L4, we have
\begin{equation}
    \bar{p}_{\rm{M}}
    = 
    \begin{cases}
        e_{2} + o(e_{2}) & \text{(for L1 and L7)}\\
        3e_{2} + o(e_{2}) & \text{(for L3 and L4)}.
    \end{cases}
\end{equation}
From Eq.~\eqref{hC}, the expected probability that a wild-type cooperates with an ALLD mutant is
\begin{equation}
    \bar{h}_{\rm{M}}^{\rm{C}}=h^{\rm{C}}(\bar{p}_{\rm{M}})=\bar{p}_{\rm{M}}+(1-\bar{p}_{\rm{M}})e_2\acp^{\rm{BB}}.
\end{equation}
Since $\acp^{\rm{BB}} = 1$ for L1, and $\acp^{\rm{BB}} = 0$ for L3, L4 and L7, we have
\begin{equation}
    \bar{h}_{\rm{M}}^{\rm{C}}
    = 
    \begin{cases}
        2e_{2} + o(e_{2}) & \text{(for L1)}\\
        3e_{2} + o(e_{2}) & \text{(for L3 and L4)}\\
        e_{2} + o(e_{2}) & \text{(for L7)}.
    \end{cases}
\end{equation}
By putting $a_{\rm{M}}=0$ in Eq.~\eqref{umutant} we obtain the expected payoff of ALLD mutants as
\begin{equation}\label{uALLDforType1}
    u_{\rm{M}}
    = 
    \begin{cases}
        2e_{2}\cdot b + o(e_{2}) & \text{(for L1)}\\
        3e_{2}\cdot b + o(e_{2}) & \text{(for L3 and L4)}\\
        e_{2}\cdot b + o(e_{2}) & \text{(for L7)}.
    \end{cases}
\end{equation}
Thus, by comparing Eq.~\eqref{uType1} and Eq.~\eqref{uALLDforType1}, wild-types are resistant to ALLD mutants if and only if
\begin{align}
    b/c>1.
\end{align}
in the limit of $e_{2} \to 0$ for all Type-1 wild-types.

\renewcommand{\thefigure}{B\arabic{figure}}
\setcounter{figure}{0}

\section{Detailed calculation}\label{App_detailed}
To solve Eq.~\eqref{phi*_p} and to obtain the equilibrium distribution $\phi^{*}(p)$, we iteratively solve the functional recursion
\begin{equation}
    \phi_{t+1}(p)
    =\int_{0}^{1}dp'\int_{0}^{1}dp''\sum_{A}h^{A}(p')\delta(p-f^{A}(p',p''))\phi_{t}(p')\phi_{t}(p''),
    \label{phit_p}
\end{equation}
where $t=0,1,2,\cdots$ is a non-negative integer, by always assuming that the function $\phi_{t}(p)$ is given in the form of Eq.~\eqref{candidate_form}; i.e., a finite sum of Dirac delta functions. If we find that $\phi_{t^{*}+1}(p) = \phi_{t^{*}}(p)$ holds at some $t^{*} \geq 0$, then it is a solution to Eq.~\eqref{phi*_p}. 

Suppose that 
\begin{equation}
    \phi_{t}(p) = \sum_{i=1}^{m_1} \kappa_{i} \delta_{1-k_{i} e_{2}} (p) + \sum_{i=1}^{m_2} \lambda_{i} \delta_{l_{i} e_{2}} (p)
\end{equation}
holds. Then, from Eq.~\eqref{phit_p} we have
\begin{equation}\label{phi_iteration}
    \begin{split}
    \phi_{t+1}(p)
        &= \left. \sum_{i=1}^{m_{1}} \sum_{j=1}^{m_{1}} \kappa_{i} \kappa_{j} \{ h^{\rm{C}}(p') \delta_{f^{\rm{C}} (p',p'')} (p) + h^{\rm{D}}(p') \delta_{f^{\rm{D}} (p',p'')} (p) \} \right|_{p'=1-k_{i} e_{2}, p''=1-k_{j} e_{2}}\\
        &+ \left. \sum_{i=1}^{m_{1}} \sum_{j=1}^{m_{2}} \kappa_{i} \lambda_{j} \{ h^{\rm{C}}(p') \delta_{f^{\rm{C}} (p',p'')} (p) + h^{\rm{D}}(p') \delta_{f^{\rm{D}} (p',p'')} (p) \} \right|_{p'=1-k_{i} e_{2}, p''=l_{j} e_{2}}\\
        &+ \left. \sum_{i=1}^{m_{2}} \sum_{j=1}^{m_{1}} \lambda_{i} \kappa_{j} \{ h^{\rm{C}}(p') \delta_{f^{\rm{C}} (p',p'')} (p) + h^{\rm{D}}(p') \delta_{f^{\rm{D}} (p',p'')} (p) \} \right|_{p'=l_{i} e_{2}, p''=1-k_{j} e_{2}}\\
        &+ \left. \sum_{i=1}^{m_{2}} \sum_{j=1}^{m_{2}} \lambda_{i} \lambda_{j} \{ h^{\rm{C}}(p') \delta_{f^{\rm{C}} (p',p'')} (p) + h^{\rm{D}}(p') \delta_{f^{\rm{D}} (p',p'')} (p) \} \right|_{p'=l_{i} e_{2}, p''=l_{j} e_{2}}.
    \end{split}
\end{equation}
Each line is further calculated, as follows.

{\bf Case 1. Interaction of a good recipient and good donor}: We consider an event that the recipient with goodness $p'=1-k_{i}e_2$ and the donor with goodness $p''=1-k_{j}e_2$ interact. This event occurs with probability $\kappa_{i} \kappa_{j}$, which can be either $O(1)$, $O(e_{2})$, or $O(e_{2}^{2})$. We consider only terms of up to $O(e_2)$ in the following. From Eq.~\eqref{hC_p}, the donor cooperates with probability
\begin{equation}
\begin{aligned}
    h^{\rm{C}}(p')&= p' + (1-p') e_{2} \acp^{\rm{BB}} \\
    &= 1-k_{i} e_2+ (k_{i} e_2) \cdot e_{2} \acp^{\rm{BB}}\\
    &=1-k_{i} e_2+o(e_2).
\end{aligned}
\end{equation}
In this case, the donor's goodness is updated to
\begin{equation}
\begin{aligned}
    f^{\rm{C}}(p',p'')&=(1-e_2)+\{ \snp^{\rm{CBB}} - (1-e_{2}) \} (1-p')(1-p'')\\
    &=(1-e_2) +\{ \snp^{\rm{CBB}} - (1-e_{2}) \} k_{i} k_{j} e_2^2 \\
    &=1-e_2+ o(e_2).
\end{aligned}
\end{equation}
On the other hand, the donor defects with probability
\begin{align}
    h^{\rm{D}}(p')=1-h^{\rm{C}}(p')=k_{i} e_2+o(e_2).
\end{align}
In this case, the donor's goodness is updated to
\begin{equation}
\begin{aligned}
    f^{\rm{D}}(p',p'')&=e_2p'+(1-e_2)(1-p')p''+\snp^{\rm{DBB}}(1-p')(1-p'')\\
    &=e_2(1-k_{i} e_2)+(1-e_2)k_{i} e_2(1-k_{j} e_2)+\snp^{\rm{DBB}}k_{i} k_{j} e_2^2\\
    &=(k_{i}+1)e_2+o(e_2).
\end{aligned}
\end{equation}
Therefore, the first line of Eq.~\eqref{phi_iteration} can be calculated as
\begin{equation}
    \sum_{i=1}^{m_{1}} \sum_{j=1}^{m_{1}} \kappa_{i} \kappa_{j} \{ (1-k_{i} e_{2}) \delta_{1-e_{2}} (p) + k_{i} e_{2} \delta_{(k_{i}+1)e_2} (p) \} + o(e_{2}).
\end{equation}

{\bf Case 2. Interaction of a good recipient and bad donor}: 
We consider an event that the recipient with goodness $p'=1-k_{i} e_2$ and the donor with goodness $p''=l_{j}e_2$ interact. This event occurs with probability $\kappa_{i} \lambda_{j}$, which can be either $O(e_{2})$ or $O(e_{2}^{2})$. We consider only terms of up to $O(1)$ in the following, because any terms of $O(e_{2})$ or higher, after the multiplication by the factor $\kappa_{i} \lambda_{j}$, become $O(e_{2}^{2})$ or higher. From Eq.~\eqref{hC_p}, the donor cooperates with probability
\begin{equation}
\begin{aligned}
    h^{\rm{C}}(p')
    &= p' + (1-p') e_{2} \acp^{\rm{BB}} \\
    &=1-k_{i} e_2+(k_{i} e_2) \cdot e_{2} \acp^{\rm{BB}}\\
    &=1 + o(1).
\end{aligned}
\end{equation}
In this case, the donor's goodness is updated to
\begin{equation}
\begin{aligned}
    f^{\rm{C}}(p',p'')
    &=(1-e_2)+\{ \snp^{\rm{CBB}} - (1-e_{2}) \} (1-p')(1-p'')\\
    &=(1-e_2)+\{ \snp^{\rm{CBB}} - (1-e_{2}) \} k_{i} e_{2} (1-l_{j} e_{2})\\
    &=1-\{(1-\snp^{\rm{CBB}})k_{i}+1\}e_2+o(e_2) \\
    &=1-\{(1-\snptil^{\rm{CBB}})k_{i}+1\}e_2+o(e_2),
\end{aligned}
\end{equation}
where $\snptil^{\rm{CBB}}$ represents the value of $\snp^{\rm{CBB}}$ evaluated at $e_{2}=0$. That is, if $\snp^{\rm{CBB}}=1-e_{2}$ then $\snptil^{\rm{CBB}}=1$, and if $\snp^{\rm{CBB}}=e_{2}$ then $\snptil^{\rm{CBB}}=0$. We do not have to discuss the event of defection because the event occurs with probability $h^{\rm{D}}(p')=1-h^{\rm{C}}(p')=o(1)$. Therefore, the second line of Eq.~\eqref{phi_iteration} can be calculated as
\begin{equation}
    \sum_{i=1}^{m_{1}} \sum_{j=1}^{m_{2}} \kappa_{i} \lambda_{j} \delta_{1-\{(1-\snptil^{\rm{CBB}})k_{i}+1\}e_2} (p) + o(e_{2}).
\end{equation}

{\bf Case 3. Interaction of a bad recipient and good donor}: We consider an event that the recipient with goodness $p'=l_{i} e_2$ and the donor with goodness $p''=1-k_{j}e_2$ interact. This event occurs with probability $\lambda_{i} \kappa_{j}$, which can be either $O(e_{2})$ or $O(e_{2}^{2})$. We consider only terms of up to $O(1)$ in the following, because any terms of $O(e_{2})$ or higher, after the multiplication by the factor $\lambda_{i} \kappa_{j}$, become $O(e_{2}^{2})$ or higher. From Eq.~\eqref{hC_p}, the donor cooperates with probability
\begin{equation}
\begin{aligned}
    h^{\rm{D}}(p')
    &= 1- \{p' + (1-p') e_{2} \acp^{\rm{BB}} \} \\
    &=1-\{ l_{i} e_2 + (1-l_{i} e_2) e_{2} \acp^{\rm{BB}}\}\\
    &=1+o(1).
\end{aligned}
\end{equation}
In this case, the donor's goodness is updated to
\begin{equation}
\begin{aligned}
    f^{\rm{D}}(p',p'')
    &=e_2 p'+(1-e_2)(1-p')p''+\snp^{\rm{DBB}}(1-p')(1-p'')\\
    &=l_{i} e_2^2+(1-e_2)(1-l_{i} e_2)(1-k_{j} e_2)+\snp^{\rm{DBB}}(1-l_{i} e_2) k_{j} e_2\\
    &=1- \{ l_{i} +(1-\snp^{\rm{DBB}})k_{j} +1 \}e_2+o(e_2) \\
    &=1- \{ l_{i} +(1-\snptil^{\rm{DBB}})k_{j} +1 \}e_2+o(e_2) \\
\end{aligned}
\end{equation}
where $\snptil^{\rm{DBB}}$ represents the value of $\snp^{\rm{DBB}}$ evaluated at $e_{2}=0$. That is, if $\snp^{\rm{DBB}}=1-e_{2}$ then $\snptil^{\rm{DBB}}=1$, and if $\snp^{\rm{DBB}}=e_{2}$ then $\snptil^{\rm{DBB}}=0$. We do not have to discuss the event of cooperation because the event occurs with probability $h^{\rm{C}}(p')=1-h^{\rm{D}}(p')=o(1)$. Therefore, the third line of Eq.~\eqref{phi_iteration} can be calculated as
\begin{equation}
    \sum_{i=1}^{m_{2}} \sum_{j=1}^{m_{1}} \lambda_{i} \kappa_{j} \delta_{1- \{ l_{i} +(1-\snptil^{\rm{DBB}})k_{j} +1 \}e_2} (p) + o(e_{2}).
\end{equation}

{\bf Case 4. Interaction of a bad recipient and bad donor}: We do not have to consider an event that the recipient with goodness $p'=l_{i} e_2$ and the donor with goodness $p''=l_{j} e_2$ interact because the event occurs with the probability of $\lambda_{i} \lambda_{j}=O(e_2^2)$. Therefore, the fourth line of Eq.~\eqref{phi_iteration} is simply $o(e_{2})$.

\begin{table}[h!]
\centering
\caption{How to calculate $\phi_{t+1}(p)$ from $\phi_{t}(p)$ for Type-1 norms.}
\begin{tabular}{ccc|ccc}
    \hline
    $p'$ & $p''$ & event scale & $A$ & $h^{\rm{A}}(p')$ & $p=f^{A}(p',p'')$ \\
    \hline
    \multirow{2}{*}{$1-k_{i} e_2$} & \multirow{2}{*}{$1-k_{j} e_2$} & \multirow{2}{*}{$O(1)$ or $O(e_2)$} & $\rm{C}$ & $1-k_{i}e_2$ & $1-e_2$ \\
    & & & $\rm{D}$ & $k_{i}e_2$ & $(k_{i}+1)e_2$ \\
    $1-k_{i} e_2$ & $l_{j} e_2$ & $O(e_2)$ & $\rm{C}$ & $1$ & $1-\{(1-\snptil^{\rm{CBB}})k_{i}+1\}e_2$ \\
    $l_{i} e_2$ & $1-k_{j} e_2$ & $O(e_2)$ & $\rm{D}$ & $1$ & $1-\{l_{i} +(1-\snptil^{\rm{DBB}})k_{j}+1\} e_2$ \\
    \hline
\end{tabular}
\label{TA03}
\end{table}

The calculations in {\bf Cases~1 to 4} are summarized in Table~\ref{TA03}. By using these results, we now solve Eq.~\eqref{phit_p} iteratively. As an initial function we choose
\begin{equation}
    \phi_{0}(p) = \delta_{1-e_{2}}(p).
\end{equation}
A direct calculation shows, up to $O(e_{2})$, that
\begin{equation}
    \begin{split}
        \phi_{1}(p) &= (1-e_{2}) \delta_{1-e_{2}}(p) + e_{2} \delta_{2e_{2}}(p)  \\
        \phi_{2}(p) &= (1-3e_{2}) \delta_{1-e_{2}}(p) + e_{2} \delta_{1-(2-\snptil^{\rm{CBB}})e_2}(p) + e_{2} \delta_{1-(4-\snptil^{\rm{DBB}})e_2}(p)+ e_{2} \delta_{2e_{2}}(p)  \\
        \phi_{3}(p) &= (1-3e_{2}) \delta_{1-e_{2}}(p) + e_{2} \delta_{1-(2-\snptil^{\rm{CBB}})e_2}(p) + e_{2} \delta_{1-(4-\snptil^{\rm{DBB}})e_2}(p)+ e_{2} \delta_{2e_{2}}(p)  \quad \left(=\phi_{2}(p)\right),
    \end{split}
\end{equation}
and therefore the solution to Eq.~\eqref{phi*_p} is
\begin{equation}\label{phi*_p_general_solution}
    \phi^{*}(p) = (1-3e_{2}) \delta_{1-e_{2}}(p) + e_{2} \delta_{1-(2-\snptil^{\rm{CBB}})e_2}(p) + e_{2} \delta_{1-(4-\snptil^{\rm{DBB}})e_2}(p)+ e_{2} \delta_{2e_{2}}(p).
\end{equation}
Substituting 
\begin{equation}
    (\snptil^{\rm{CBB}}, \snptil^{\rm{DBB}}) =
    \begin{cases}
        (1,0) & \text{(for L1)}\\
        (1,1) & \text{(for L3)}\\
        (0,1) & \text{(for L4)}\\
        (0,0) & \text{(for L7)}
    \end{cases}
\end{equation}
in Eq.~\eqref{phi*_p_general_solution} gives us Eq.~\eqref{phi*_solution}.

\end{document}